\documentclass[journal]{IEEEtran}
\ifCLASSINFOpdf
  \usepackage[pdftex]{graphicx}
  \graphicspath{{../pdf/}{../jpeg/}}
  \DeclareGraphicsExtensions{.pdf,.jpeg,.png}
\else
  \usepackage[dvips]{graphicx}
  \graphicspath{{../eps/}}
  \DeclareGraphicsExtensions{.eps}
\fi
\usepackage{epsfig}
\usepackage{booktabs}
\usepackage{epstopdf}
\usepackage{amssymb}
\usepackage{mathtools}
\usepackage{subfigure}
\usepackage{algorithmic}

\usepackage{array}
 \usepackage{mathrsfs}

\usepackage{amssymb}
\usepackage{amsmath}
\usepackage{cite}
\usepackage{url}
\usepackage{xcolor}
\usepackage{cite,graphicx,amsmath,amssymb}
\usepackage{subfigure}
\usepackage{psfrag}
\usepackage{pstool}
\usepackage{fancyhdr}
\usepackage{mdwmath}
\usepackage{mdwtab}
\usepackage{amsthm}
\usepackage{suffix}
\usepackage{mathtools}
\allowdisplaybreaks[4]
\usepackage{xpatch}
\usepackage{multirow}

\xpatchbibmacro{name:andothers}{%
  \bibstring{andothers}%
}{%
  \bibstring[\emph]{andothers}%
}{}{}

\DeclarePairedDelimiterX\MeijerM[3]{\lparen}{\rparen}%
{\begin{smallmatrix}#1 \\ #2\end{smallmatrix}\delimsize\vert\,#3}

\newcommand\MeijerG[8][]{%
  G^{\,#2,#3}_{#4,#5}\MeijerM[#1]{#6}{#7}{#8}}

\WithSuffix\newcommand\MeijerG*[7]{%
  G^{\,#1,#2}_{#3,#4}\MeijerM*{#5}{#6}{#7}}

\newtheorem{theorem}{Theorem}

\newtheorem{lemma}{Lemma}

\newtheorem{corollary}{Corollary}

\begin{document}
\title{On NOMA-Based mmWave Communications
\thanks{Manuscript received**, 2019; revised **, 2019; accepted **, 2019. The associate editor coordinating the review of this paper and approving it for publication was ***. }
\thanks{The authors are with Computer, Electrical and Mathematical Science and Engineering Division, King Abdullah University of Science and Technology (KAUST), Thuwal 23955-6900, Saudi Arabia.}
}

\author{Yu Tian,  Gaofeng Pan,~\IEEEmembership{Senior Member,~IEEE}, and Mohamed-Slim Alouini,~\IEEEmembership{Fellow,~IEEE}}

\maketitle
\begin{abstract}
    Non-orthogonal multiple access (NOMA) and millimeter-wave (mmWave) communication are two promising techniques to increase the system capacity in the fifth-generation (5G) mobile network. The former can achieve high spectral efficiency by modulating the information in power domain and the latter can provide extremely large spectrum resources. Fluctuating two-ray (FTR) channel model has already been proved to accurately agree with the small-scale fading effects in mmWave communications in experiments. In this paper, the performance of NOMA-based communications over FTR channels in mmWave communication systems is investigated in terms of outage probability (OP) and ergodic capacity (EC). Specifically, we consider the scenario that one base station (BS) transmits signals to two users simultaneously under NOMA scheme. The BS and users are all equipped with a single antenna. Two power allocation strategies are considered: the first one is a general (fixed) power allocation scheme under which we derive the OP and EC of NOMA users in closed form; the other one is an optimal power allocation scheme that can achieve the maximum sum rate for the whole system. Under the second scheme, not only the closed-form OP and EC but also the upper and lower bounds of EC are derived. Furthermore, we also derive the asymptotic expression for the OP in high average SNR region to investigate the diversity order under these two schemes. Finally, we show the correctness and accuracy of our derived expressions by Monte-Carlo simulation.
\end{abstract}


\section{Introduction}
Non-orthogonal multipath access (NOMA) and mmWave communication are two popular techniques in the fifth-generation (5G) mobile networks to solve the serious issues of explosively increasing bandwidth demands in current communication systems\cite{zhang2017capacity,sun2018performance,li2019performance,zhang2017non,li2019performances, zhang2018enhancing,makki2020survey}. Under NOMA scheme, the transmitted signals are modulated in the power domain along with the time, frequency, and code domains, which will significantly improve the spectrum efficiency. MmWave communication operates from 30 to 300 GHz, which occupies a very large bandwidth, and thus it can efficiently increase the system capacity. 

NOMA scheme implements power modulation via allocating higher power to the signals transmitted to the users with worse channel conditions. At the users' side, the receivers with better channel conditions decode the signals through a technique called successive interference cancellation (SIC) while these receivers with worse conditions decode their information directly. As NOMA adds another freedom of power modulation, it can significantly improve the spectrum efficiency, user fairness, and flexibility compared with traditional orthogonal multiple access.
Many works have been conducted to analyze the performance of NOMA schemes in various communication systems. In \cite{xiang2020secure} and \cite{xiang2019physical}, connection outage probability (OP), secrecy OP, and effective secrecy throughput were investigated in hybrid automatic repeat request-assisted NOMA networks and cognitive radio inspired NOMA networks. To improve the system performance, a relay was introduced in the cooperative NOMA network and the OP was studied in \cite{xu2018optimal}. In \cite{lei2018secrecy}, a multiple-input multiple-output (MIMO) NOMA network was considered and secrecy performance was studied under a max-min transmit antenna selection strategy. In \cite{xu2016outage}, 1-bit feedback was utilized in NOMA systems and the outage performance of the downlink was investigated. 

Due to the outstanding benefits offered by mmWave communications, some other researchers have investigated on the performance of NOMA-based mmWave communication systems. In \cite{zhang2017capacity}, capacity analysis was conducted for integrated NOMA mmWave-massive-MIMO systems over a uniform random single-path mmWave channel model extended by the angle of arrival. In \cite{sun2018performance}, the OP of downlink NOMA in multi-cell mmWave networks was studied over Nakagami-$m$ fading. In \cite{li2019performance}, coverage probability and sum rate of agile-beam NOMA were investigated in mmWave networks over single-path channels. In \cite{zhang2017non}, the coverage probability, the average number of served users, and the sum rate of NOMA were studied 
for the cooperative multicast mmWave wireless networks over Nakagami-$m$ fading channels. In \cite{li2019performances}, the OP and the ergodic capacity (EC) were investigated in downlink MIMO-NOMA mmWave cellular networks with D2D communications  over Nakagami-$m$ fading channels. However, the small-scale fading effects in these considered mmWave communication scenarios were modeled by using Nakagami-$m$ fading or some other simplified fadings.

On the other hand, fluctuating two-ray (FTR) fading distribution has been proposed and proved that it can provide a better fit for the small-scale fading characteristics in mmWave communications \cite{romero2016fluctuating,romero2017fluctuating}. Compared with other fading models, including generalized two-ray model proposed in \cite{rao2015mgf}, it introduces the fluctuating amplitudes of specular components to replace the constant ones in previous small-scale fading models. Therefore, it can well capture the wide heterogeneity of random amplitude fluctuations caused by multiple scatters in mmWave device-to-device propagation environments and holds a wide generality. It has been experimentally verified that FTR model can perform better in mmWave frequency range than Rician, Rayleigh, and Nakagami-$m$ fading models, which are the special cases of FTR fading. Moreover, it can also well generalize the two-wave with diffuse power (TWDP) model and address the issue that TWDP model does not have the closed-form expression\cite{zhao2019secure}.  \cite{zhang2017new,zhao2018different} improve the expressions in \cite{romero2017fluctuating} and give more generalized probability density function (PDF) and cumulative distribution function (CDF) which are easier for calculation and suitable for any positive real parameter $m$. In \cite{zhao2018different,zhao2019secure,zheng2019wireless,bilim2019average}, the authors worked on the performance of several power adaption methods, cognitive radio networks, UAV relay communications, and QAM modulation over FTR channels. However, so far, the performance of NOMA scheme in mmWave communications over FTR channels has not been studied yet. 

Motivated by this observation, in this work, a downlink NOMA system over FTR channels is considered, where a base station (BS) sends data to two users. Especially, one user is closer to the BS than the other one. The BS and two users are all equipped with a single antenna.

The main contributions of this paper are summarized as follows:
\begin{enumerate}
\item[1)]
We derive the closed-form expressions for the OP of these two users and EC of the considered system under a general (fixed) power allocation (GPA) scheme, as well as the asymptotic expressions for OP which reveal the diversity orders of these two users;
\item[2)]
We investigate the EC and OP of the considered system under an optimal power allocation (OPA) scheme by deriving the closed-form and asymptotic expressions for OP, and the closed-form expression, upper bound, and lower bound for EC;
\item[3)]
We systematically study and summarize the impacts of the parameters of FTR channels on the OP and EC of the considered system under both GPA and OPA schemes.
\end{enumerate}

 The rest of this paper is organized as follows. Section \ref{systmemodel} describes the considered NOMA-based wireless communication system over FTR channels. Section \ref{generalscheme} conducts the analysis of the OP and EC of the considered system under GPA scheme by deriving the closed-form and asymptotic expressions of OP, and the closed-form of EC. Section \ref{optimalscheme} investigates the closed-form and asymptotic expressions of OP, and the closed-form expression, upper bound, and lower bound of EC under OPA scheme. Section \ref{discusionGPAOPA} gives some discussions on GPA and OPA schemes, and some remarks are given. Section \ref{simulations} studies the impacts of FTR channel parameters on OP and EC. Also, numerical results are presented and discussed. Finally, Section \ref{conclusion} concludes the paper.

\section{System Model}\label{systmemodel}
In this work, we consider the scenario that two users ($p$ and $q$) are served by a BS but have different distances from it, while user $p$ is closer to the BS than user $q$. According to NOMA scheme, the BS simultaneously broadcasts the superposed signal $x$ containing the information for both $p$ and $q$ with different transmitted powers. Supposing that the fraction of the transmit power allocated to user $p$ is $a$, user $q$ will occupy $1-a$ fraction of the power. The signal received by user $t\ (t\in\{p,q\})$ is defined as 
\begin{align}
    y_t = \sqrt{h_tQ_t} x+n_t,
\end{align}
where $h_t$ is the FTR channel power gain, $Q_t$ is the multiplication of other channel gains (antenna gains and path loss, etc.), and $n_t$ is the additive white Gaussian noise with variance $N_0$.

As user $p$ is closer to the BS than user $q$, $Q_p> Q_q$ (usually $Q_p\gg Q_q$ \cite{ding2015impact}) can be obtained. According to NOMA scheme, the transmitted signal for the further user $q$ occupies more fraction of the power, which means $a<0.5$. At the receiver, user $q$ decodes its information directly. According to SIC, user $p$ firstly decodes user $q$'s information, then decodes its own information after subtracting these already decoded. Therefore, the signal-to-interference-plus-noise ratios (SINRs) of users $p$ and $q$ are shown as
\begin{align}\label{snrp}
    \gamma_p=a \overline{\gamma}Q_ph_p
\end{align}
and 
\begin{align}\label{snrq}
    \gamma_q=\frac{(1-a)\overline{\gamma}Q_qh_q}{a \overline{\gamma}Q_qh_q+1},
\end{align}
where $\overline{\gamma}$ is the average SNR of the signal broadcast by the BS. 

The FTR channel model is composed of a diffuse component and two specular components with random phases which cause fluctuating amplitudes \cite{romero2017fluctuating}. Assuming that user $p$ and $q$ experience independent FTR fading, the PDF of $h_t\ (t\in\{p,q\})$ is given by \cite{zhao2019secure}
\begin{align}
    f_{h_t}=\sum_{j_t=0}^{\infty}H_t\frac{x^{j_t}}{j_t!(2\sigma_t^2)^{j_t+1}}\exp\left(-\frac{x}{2\sigma_t^2}\right),
\label{PDF}
\end{align}
where $H_t=\frac{m_t^{m_t}K_t^{j_t}d_{j_t}}{\Gamma(m_t)j_t!}$, $\Gamma(\cdot)$ represents Gamma function, $m_t$ is one parameter in the PDF of Gamma distribution with unit mean, $K_t$ is the ratio of the average power of the dominant waves and that of other diffuse multipath waves, $\sigma^2_t$ is the variance of the real part of the diffuse signal, $d_{j_t}$ is defined as
\begin{align}
    d_{j_t} &= \sum_{k=0}^{j_t}\binom{j_t}{k}\left(\frac{\Delta_t}{2}\right)^k\sum_{l=0}^{k}\binom{k}{l}\Gamma(j_t+m_t+2l-k)\notag\\
    &~~~\times\frac{\exp{(\frac{\pi(2l-k)i}{2})P_{j_t+m_t-1}^{k-2l}}\left(\frac{m_t+K_t}{\sqrt{(m_t+K_t)^2-(K_t\Delta_t)^2}}\right)}{\left(\sqrt{(m_t+K_t)^2-(K_t\Delta_t)^2}\right)^{j_t+m_t}},
\end{align}
where $\binom{\cdot}{\cdot}$ defines the binomial coefficient, $i$ is the imaginary unit, $\Delta_t$ characterizes two dominant wave powers' relations, and $P(\cdot)$ represents the first kind Legendre function \cite{gradshteyn2007}.

Accordingly, the CDF of $ h_t$ can be given as
\begin{align}
    F_{h_t}(x)=1-\sum_{j_t=0}^{\infty}H_t\exp\left(-\frac{x}{2\sigma_t^2}\right)\sum_{n_t=0}^{j_t}\frac{\left(x/(2\sigma_t^2)\right)^{n_t}}{n_t!}.
\label{CDF}
\end{align}

\section{Performance analysis under GPA scheme}\label{generalscheme}

\subsection{Outage Analysis}
OP is the probability that the SINR of the user \(t\), $\gamma_t\ (t\in\{p,q\})$, falls below a given SINR threshold $\gamma_{\rm th}$\cite{yan2018performance}, which can be given as
\begin{align}
    OP_t(\gamma_{\rm th})={\rm Pr}\{\gamma_t \leq \gamma_{\rm th}\},
\label{op}
\end{align}
\subsubsection{The OP of User $p$}
Substituting \eqref{snrp} and \eqref{CDF} into \eqref{op}, the OP of user $p$ is calculated as 
{\begin{align}\label{oppgeneral}
    OP_p(\gamma_{\rm th})
    &={\rm Pr}\left\{h_p\leq\frac{\gamma_{\rm th}}{a \overline{\gamma} Q_p} \right\}\notag\\
    &=F_{h_p}\left(\frac{\gamma_{\rm th}}{a \overline{\gamma} Q_p}\right)\notag\\
    &=1-\sum_{j_p=0}^{\infty}H_p\exp\left(-\frac{\alpha \gamma_{\rm th}}{a}\right)\notag\\
    &~~~\times\sum_{n_p=0}^{j_p}\frac{1}{n_p!}\left(\frac{\alpha \gamma_{\rm th}}{a}\right)^{n_p},
\end{align}}
where $\alpha=\frac{1}{2\sigma_p^2\overline{\gamma}Q_p}$.

\subsubsection{The OP of User $q$}
Employing \eqref{snrq}, \eqref{CDF} and \eqref{op}, the OP of user $q$ is presented as 
{
\begin{align}\label{opqgeneral}
    OP_q(\gamma_{\rm th})
    &={\rm Pr}\left\{a_{\rm th}h_q\leq\frac{\gamma_{\rm th} }{  \overline{\gamma}Q_q}\right\}\notag\\
    &=\left\{\begin{array}{ll}
     {\rm Pr}\left\{h_q\leq\frac{\gamma_{\rm th} }{  a_{\rm th}\overline{\gamma}Q_q}\right\},   & {\rm{if}}~a_{\rm th}>0; \\
      1, & {\rm{if}}~a_{\rm th}=0;\\
     {\rm Pr}\left\{h_q\ge\frac{\gamma_{\rm th} }{  a_{\rm th}\overline{\gamma}Q_q}\right\},
       & {\rm{otherwise}}
    \end{array}
    \right.\notag\\
    &=\left\{\begin{array}{ll}
       1,  & {\rm{if}}~0\ge a_{\rm th};       \\
       F_{h_q}\left(\frac{\gamma_{\rm th} }{ a_{\rm th} \overline{\gamma}Q_q}\right),   & {\rm{otherwise}}
    \end{array}
    \right.\notag\\
    &=\left\{\begin{array}{ll}
    1,  & {\rm{if}}~0\ge a_{\rm th}; \\
      \begin{array}{@{}ll@{}}
      1-\sum_{j_q=0}^{\infty}H_q  \\ \times\exp{\left(-\frac{\beta \gamma_{\rm th}}{a_{\rm th} }\right)}\\
    \times\sum_{n_q=0}^{j_q}\frac{1}{n_q!}\left(\frac{\beta \gamma_{\rm th}}{a_{\rm th} }\right)^{n_q} \end{array}& {\rm{otherwise}}
    \end{array}
    \right.,
\end{align}
where $\beta=\frac{1}{2\sigma_q^2\overline{\gamma}Q_q}$, $a_{\rm th}=1-a-a\gamma_{\rm th}$, and $h_q$ always has non-negative values.}

\subsubsection{Asymptotic OP at High SNR}
When $\overline{\gamma}\gg0$, the asymptotic CDF of $ h_t$  is given by \cite[Eq. (18)]{zhao2019secure} as
\begin{align}
    F_{ h_t}^{\infty}(x) \approx \frac{m_t^{m_t}d_{j_t=0}x}{2\sigma_t^2\Gamma(m_t)},
\label{highsnrcdf}
\end{align}
where $d_{j_d=0}$ denotes the value of $d_{j_d}$ when $j_d=0$.

According to \eqref{oppgeneral} and \eqref{highsnrcdf}, user $p$'s asymptotic OP is expressed as
\begin{align}\label{asympoppgeneral}
    OP_p^{\infty}(\gamma_{\rm th})&=F^{\infty}_{h_p}\left(\frac{\gamma_{\rm th}}{a \overline{\gamma} Q_p}\right)=\frac{m_p^{m_p}d_{j_p=0}\gamma_{\rm th}}{2a\sigma_p^2 Q_p\Gamma(m_p)}\overline{\gamma}^{-1},
\end{align}
which shows that the diversity order of user $p$ is 1.

Similarly, user $q$'s asymptotic OP is defined as
\begin{align}\label{aympopqgeneral}
     OP_q^{\infty}(\gamma_{\rm th})     =&\left\{\begin{array}{ll}
          F^{\infty}_{h_q}\left(\frac{\gamma_{\rm th} }{ a_{\rm th}\overline{\gamma}Q_q}\right), & {\rm{if}}~a_{\rm th}>0; \\
          1, & {\rm{otherwise}} 
     \end{array}\right.\notag\\
    =&\left\{\begin{array}{ll}
        \frac{m_q^{m_q}d_{j_q=0}\gamma_{\rm th}}{2a_{\rm th}\sigma_q^2 Q_q\Gamma(m_q)}\overline{\gamma}^{-1}, & {\rm{if}}~a_{\rm th}>0; \\
        \overline{\gamma}^{0}, & {\rm{otherwise}}
    \end{array}\right.,
\end{align}
which implies that the diversity order of user $q$  has two values: 1 when $a+a\gamma_{\rm th}<1$ and 0 when $a+a\gamma_{\rm th}\ge1$.

\subsection{Ergodic Capacity}
The EC considered in this paper is defined as 
$C_{\rm erg}={\mathbb{E}}\left[\log_2(1+\gamma_p)\right]+{\mathbb{E}}\left[\log_2(1+\gamma_q)\right]$
and gives the average information transmission rate of the whole NOMA system\cite{yan2018performance}. Exploiting \eqref{snrp} and \eqref{snrq}, we can get $C_{\rm erg}$ in \eqref{ECnormal}, demonstrated on the top of next page.
\begin{figure*}
\begin{align}\label{ECnormal}
    C_{\rm erg}&={\mathbb{E}}\left[\log_2(1+a \overline{\gamma}Q_ph_p)\right]+{\mathbb{E}}\left[\log_2\left(1+\frac{(1-a)\overline{\gamma}Q_qh_q}{a \overline{\gamma}Q_qh_q+1}\right)\right]\notag\\
    &=\int_0^{\infty}\log_2\left(1+a \overline{\gamma}Q_px\right)f_{h_p}(x)\mathrm{d}x+\int_0^{\infty}\log_2\left(\frac{1+\overline{\gamma}Q_qy}{a \overline{\gamma}Q_qy+1}\right)f_{h_q}(y)\mathrm{d}y\notag\\
    &=\underbrace{\int_0^{\infty}\log_2\left(1+a \overline{\gamma}Q_px\right)f_{h_p}(x)\mathrm{d}x}_{I_0}+\underbrace{\int_0^{\infty}\log_2\left(1+\overline{\gamma}Q_qy\right)f_{h_q}(y)\mathrm{d}y}_{I_1}-\underbrace{\int_0^{\infty}\log_2\left(1+a \overline{\gamma}Q_qy\right)f_{h_q}(y)\mathrm{d}y}_{I_2}
\end{align}
\rule{18cm}{0.01cm}
\end{figure*}

From \eqref{ECnormal}, we can see that $I_0$, $I_1$ and $I_2$ have the same form, which is
\begin{align}\label{ecnorm}
    \Lambda(b,t)=\int_0^{\infty}\log_2\left(1+bx\right)f_{h_t}(x)\mathrm{d}x,
\end{align}
where $b\in\{a \overline{\gamma}Q_p$, $\overline{\gamma}Q_q$, $a \overline{\gamma}Q_q\}$ and $t\in\{p, q\}$.

{
To make integral in \eqref{ecnorm} tractable, we first represent $\log_2\left(1+bx\right)$ as $\log_2(1+bx)=\frac{1}{\ln2}\MeijerG*{1}{2}{2}{2}{1,1}{1,0}{bx}$ according to \cite[Eq. (11)]{adamchik1990algorithm}, where $G^{m,n}_{p,q}(\cdot)$ is the Meijer-$G$ function.
}

Employing \eqref{PDF} and \cite[Eq. (2.6.2)]{mathai1978}, \eqref{ecnorm} can be written as
\begin{align}\label{logint}
    \Lambda (b,t)  
    &=\frac{1}{\ln2}\sum_{j_t=0}^{\infty}\frac{H_t}{j_t!(2\sigma_t^2)^{j_t+1}}\notag\\
    &~~~~\times\int_{0}^{\infty}\exp\left(-\frac{x}{2\sigma_t^2}\right)x^{j_t}\MeijerG*{1}{2}{2}{2}{1,1}{1,0}{bx}\mathrm{d}x\notag\\
    &=\frac{1}{\ln2}\sum_{j_t=0}^{\infty}\frac{H_t}{j_t!}\MeijerG*{1}{3}{3}{2}{-j_t,1,1}{1,0}{2b\sigma_t^2}.
\end{align}

Based on \eqref{logint}, one can get $C_{\rm erg}$ in \eqref{cerg}, presented on the top of next page.
\begin{figure*}
\begin{align}\label{cerg}
    C_{\rm erg}=I_0+I_1-I_2=\frac{1}{\ln2}\sum_{j_p=0}^{\infty}\frac{H_p}{j_p!}\MeijerG*{1}{3}{3}{2}{-j_p,1,1}{1,0}{\frac{a}{\alpha}}+\frac{1}{\ln2}\sum_{j_q=0}^{\infty}\frac{H_q}{j_q!}\left(\MeijerG*{1}{3}{3}{2}{-j_q,1,1}{1,0}{\frac{1}{\beta}}-\MeijerG*{1}{3}{3}{2}{-j_q,1,1}{1,0}{\frac{a}{\beta}}\right)
\end{align}
\rule{18cm}{0.01cm}
\end{figure*}

\section{Performance analysis under OPA scheme}\label{optimalscheme}

{According to \eqref{snrp} and \eqref{snrq}, the sum rate of the considered system can be calculated as
\begin{align}\label{sumrate}
    R_{sum}&=R_p^{NOMA}+R_q^{NOMA}\notag\\
    &=\log_2\left(1+a\overline{\gamma}Q_ph_p\right)+\log_2\left(1+\frac{(1-a)\overline{\gamma}Q_qh_q}{a \overline{\gamma}Q_qh_q+1}\right).
\end{align}}

As pointed out in \cite{ding2016impact}, NOMA can be regarded as a special case of cognitive radio systems. To guarantee that both users $p$ and $q$ can be considered as a primary user, the transmission rates of these two users under NOMA scheme should be always greater than those under time division multiple access (TDMA) scheme \cite{yang2016general}. {The range of $a$ can be constrained as
\begin{align}
    \frac{1}{\sqrt{1+\overline{\gamma}Q_ph_p}+1}\leq a \leq \frac{1}{\sqrt{1+\overline{\gamma}Q_qh_q}+1}.
\end{align}}
{The derivation of the range of $a$ is shown in Appendix I.}

{In Appendix I, we also prove that the first derivative of $R_{sum}$ in \eqref{sumrate} with respect to $a$ is almost strictly positive. Therefore, the maximum sum rate for the whole system can be achieved when $a=\frac{1}{\sqrt{1+\overline{\gamma}Q_qh_q}+1}$.} 

Thus, the received SINRs at user $p$ and $q$ are given by

\begin{align}
    \gamma_p = \frac{\overline{\gamma} Q_p h_p}{\sqrt{1+\overline{\gamma} Q_q h_q}+1}
    \label{rp}
\end{align}
and
\begin{align}
    \gamma_q = \sqrt{1+\overline{\gamma}Q_q h_q}-1.
    \label{rq}
\end{align}

\noindent 

{Moreover, by observing that the transmit power allocation for each user under OPA scheme is different from that under GPA scheme, we have also demonstrated the feasibility of the SIC at user $p$ under OPA scheme in Appendix II.}

\subsection{Outage Analysis}

\subsubsection{The OP of User p}

Substituting \eqref{CDF} and \eqref{rp} into \eqref{op}, the OP of user $p$ is calculated in \eqref{OPp11}, described on the top of next page.

\begin{figure*}
    \begin{align}\label{OPp11}
         OP_p(\gamma_{\rm th})&={\rm Pr}\left\{ h_p \leq \frac{\gamma_{\rm th}\left(\sqrt{1+\overline{\gamma}Q_q h_q}+1\right)}{\overline{\gamma}Q_p}\right\}\notag\\
         &=\int_0^{\infty}F_{ h_p}\left(\frac{\gamma_{\rm th}\left(\sqrt{1+\overline{\gamma}Q_qy}+1\right)}{\overline{\gamma}Q_p}\right)f_{ h_q}(y)\mathrm{d}y\notag\\
        &=\int_0^{\infty}f_{h_q}(y)\mathrm{d}y-\int_0^{\infty}\sum_{j_p=0}^{\infty}H_pe^{-\gamma_{\rm th}\alpha-\gamma_{\rm th}\alpha\sqrt{1+\overline{\gamma}Q_qy}} \sum_{n_p=0}^{j_p}\frac{\left({\gamma_{\rm th}\alpha\left(\sqrt{1+\overline{\gamma}Q_qy}+1\right)}\right)^{n_p}}{n_p!}f_{h_q}(y)\mathrm{d}y\notag\\
        &=1-\sum_{j_p=0}^{\infty}H_pe^{-\gamma_{\rm th}\alpha} \sum_{n_p=0}^{j_p}\frac{(\gamma_{\rm th}\alpha)^{n_p}}{n_p!}\sum_{j_q=0}^{\infty}\frac{H_q}{j_q!(2\sigma_q^2)^{j_q+1}}\underbrace{\int_0^{\infty}e^{-\gamma_{\rm th}\alpha\sqrt{1+\overline{\gamma}Q_qy}-\frac{y}{2\sigma_q^2}}y^{j_q} \left(\sqrt{1+\overline{\gamma}Q_qy}+1\right)^{n_p}\mathrm{d}y}_{I_3}
    \end{align}
    \rule{18cm}{0.01cm}
\end{figure*}

Let $z=\sqrt{1+\overline{\gamma}Q_qy}$, then $y=\frac{z^2-1}{\overline{\gamma}Q_q}$. Therefore, $I_3$ in \eqref{OPp11} can be expressed as
\begin{align}
       I_3=2\exp{(\beta)}({\overline{\gamma}Q_q})^{-j_q-1}\cdot I_4.
\label{OPp2}
\end{align}
where $I_4=\int_1^{\infty}\exp{(-\alpha\gamma_{\rm th}z-\beta z^2)}\left({z-1}\right)^{j_q}\left(z+1\right)^{n_p+j_q}$ $z\mathrm{d}z$.

According to Taylor series in \cite[Eq. (1.111)]{gradshteyn2007}, $(z-1)^{j_q}$ and $(z+1)^{n_p+j_q}$ can be rewritten as
\begin{align}
  (z-1)^{j_q}=\sum\limits_{n=0}^{j_q}\binom{j_q}{n}(-1)^{j_q-n}z^n  
  \label{zminus1}
\end{align}
and
 \begin{align}
     (z+1)^{n_p+j_q}=\sum\limits_{m=0}^{n_p+j_q}\binom{n_p+j_q}{m}z^m.
     \label{zadd1}
 \end{align}

Observing \eqref{OPp2} and substituting \eqref{zminus1} and \eqref{zadd1} into the integral part $I_4$ in \eqref{OPp2}, we get
\begin{align}
    I_4=&\sum_{n=0}^{j_q}\binom{j_q}{n}(-1)^{j_q-n}\sum_{m=0}^{n_p+j_q}\binom{n_p+j_q}{m}\notag\\
    &\times \underbrace{\int_1^{\infty}\exp{\left(-\gamma_{\rm th}\alpha z-\beta z^2\right)}z^{m+n+1}\mathrm{d}z}_{I_5},
    \label{OPp3}
\end{align}
where $I_5$ can be written as
\begin{align}
    I_5=I_6-I_7,
\label{OPp4}
\end{align}
where \(I_6=\int_0^{\infty}\exp{(-\gamma_{\rm th}\alpha z-\beta z^2)}z^{m+n+1}\mathrm{d}z\) and
\(I_7=\int_0^{1}\exp{(-\gamma_{\rm th}\alpha z-\beta z^2)}z^{m+n+1}\mathrm{d}z\).

Using \cite[Eq. (3.462-1)]{gradshteyn2007}, $I_6$ can be expressed as
\begin{align}
    I_6=\nu!\left(2\beta\right)^{\frac{-\nu-1}{2}}\exp{\left(\frac{\gamma_{\rm th}^2\alpha^2}{8\beta}\right)}D_{-\nu-1}\left(\frac{\alpha\gamma_{\rm th}}{\sqrt{2\beta}}\right),
\label{OPp5}
\end{align}
where $D_{v}(\cdot)$ is the parabolic-cylinder function and $\nu=m+n+1$.

By setting $z=2t-1$ and using Chebyshev-Gauss quadrature  in  the  first case\cite{abramowitz+stegun}, which is given as $\int_{-1}^1\frac{f(x)}{\sqrt{1-x^2}}\mathrm{d}x \approx \sum\limits_{k=1}^{I} f(x_k)\pi/I$ with $x_k=\cos{\left(\frac{2k-1}{2I}\pi\right)}$, $I_7$ can be expressed as
\begin{align}
    I_7&=2^{-\nu-1}\exp{\left(-\frac{\alpha\gamma_{\rm th}}{2}-\frac{\beta}{4}\right)}\notag\\
    &~~~\times \int_{-1}^1\exp{\left(-\frac{(\alpha\gamma_{\rm th}+\beta)t}{2}-\frac{\beta}{4} t^2\right)}(1+t)^{\nu}\mathrm{d}t\notag\\
    &=2^{-\nu-1}\exp{\left(-\frac{\alpha\gamma_{\rm th}}{2}-\frac{\beta}{4}\right)}\sum_{k=1}^{I}\frac{\pi}{I}\sqrt{1-\phi_k^2}\notag\\
    &~~~\times \exp{\left(-\frac{(\alpha\gamma_{\rm th}+\beta)\phi_k}{2}-{\frac{\beta}{4}\phi_k^2}\right)}(1+\phi_k)^{\nu},
\label{OPp6}
\end{align}
where $\phi_k=\cos{\left(\frac{2k-1}{2I}\pi\right)}$.

Using \eqref{OPp11}, \eqref{OPp2}, \eqref{OPp3}, \eqref{OPp4}, \eqref{OPp5}, and \eqref{OPp6}, we can have the closed-form expression of $OP_p(\gamma_{\rm th})$ in \eqref{OPp8}, shown on the top of next page.
\begin{figure*}
\begin{align}\label{OPp8}
        OP_p(\gamma_{\rm th})=&1-2\exp{\left(-\gamma_{\rm th}\alpha+\beta\right)}\sum_{j_p=0}^{\infty}H_p \sum_{n_p=0}^{j_p}\frac{{(\gamma_{\rm th}\alpha)}^{n_p}}{n_p!}\sum_{j_q=0}^{\infty}\frac{H_q\beta^{j_q+1}}{j_q!} \sum_{n=0}^{j_q}\binom{j_q}{n}(-1)^{j_q-n}\sum_{m=0}^{n_p+j_q}\binom{n_p+j_q}{m} \notag\\
        &\times\Bigg\{\nu! \left(2\beta\right)^{\frac{-\nu-1}{2}}e^{\frac{\gamma_{\rm th}^2\alpha^2}{8\beta}}D_{-\nu-1}\left(\frac{\gamma_{\rm th}\alpha}{\sqrt{2\beta}}\right)- 2^{-\nu-1}e^{-\frac{\alpha\gamma_{\rm th}}{2}-\frac{\beta}{4}}\sum_{k=1}^{I}\frac{\pi}{I}\sqrt{1-\phi_k^2}(1+\phi_k)^{\nu} e^{-\frac{(\alpha\gamma_{\rm th}+\beta)\phi_k}{2}-{\frac{\beta}{4}\phi_k^2}}\Bigg\}
\end{align}
\rule{18cm}{0.01cm}
\end{figure*}

\subsubsection{The OP of User q}
Substituting \eqref{CDF} and \eqref{rq} into \eqref{op}, the OP of user $q$ can be given as
\begin{align}
    OP_q(\gamma_{\rm th})&={\rm Pr}\left\{ h_q \leq \frac{\gamma_{\rm th}^2+2\gamma_{\rm th}}{\overline{\gamma}Q_q}\right\}\notag\\
    &=F_{ h_q}\left(\frac{\gamma_{\rm th}^2+2\gamma_{\rm th}}{\overline{\gamma}Q_q}\right)\notag\\
    &=1-\sum_{j_q=0}^{\infty}H_q\exp\left(-\beta(\gamma_{\rm th}^2+2\gamma_{\rm th})\right)\notag\\
    &~~~\times\sum_{n_q=0}^{j_q}\frac{\left(\beta(\gamma_{\rm th}^2+2\gamma_{\rm th})\right)^{n_q}}{n_q!}.
\label{opofq}
\end{align}

\subsubsection{Asymptotic OP at High SNR}
\paragraph{Asymptotic OP of user $p$}
When $\overline{\gamma}\gg0$, $\sqrt{1+\overline{\gamma} Q_q h_q}+1\approx\sqrt{\overline{\gamma} Q_q h_q}$. According to \eqref{rp}, we have
\begin{align}
    \gamma^{\infty}_p\approx\sqrt{\overline{\gamma}}\frac{ Q_p h_p}{ \sqrt{Q_q h_q}}
    \label{inftyrp}.
\end{align}

Substituting \eqref{inftyrp}, \eqref{highsnrcdf}, and \eqref{PDF} into \eqref{op} and \eqref{OPp11} and then using \cite[Eq. (3.371)]{gradshteyn2007}, we can get the asymptotic OP of user $p$ as

\begin{align}
        OP_p^{\infty}(\gamma_{\rm th})&\approx \int_0^{\infty}F^{\infty}_{ h_p}\left(\frac{\gamma_{\rm th}\sqrt{Q_qy}}{\sqrt{\overline{\gamma}}Q_p}\right)f_{ h_q}(y)\mathrm{d}y\notag\\ &=\frac{m_p^{m_p}d_{j_p=0}\gamma_{\rm th}\sqrt{Q_q}}{2\sigma^2_pQ_p\Gamma(m_p)\sqrt{\overline{\gamma}}}\sum_{j_q=0}^{\infty}\frac{H_q}{j_q!(2\sigma_q^2)^{j_q+1}}\notag\\
        &~~~\times 
        \int_0^{\infty}y^{j_q+\frac{1}{2}}\exp\left(-\frac{y}{2\sigma_q^2}\right)\mathrm{d}y\notag\\
        &=\frac{m_p^{m_p}d_{j_p=0}\gamma_{\rm th}\sqrt{\sigma_q^2Q_q\pi}}{\sigma^2_pQ_p\Gamma(m_p)\sqrt{2}}\sum_{j_q=0}^{\infty}\frac{(2j_q+1)!!H_q}{2^{j_q+1}j_q!}\overline{\gamma}^{-\frac{1}{2}},
\label{highsnrcdfp}
\end{align}
where $(2j_q+1)!!=1\times3\times5\times\dots\times(2j_q+1)$. It shows that the diversity order of user $p$ is 0.5.
\paragraph{Asymptotic OP of user q}
Similarly, substituting  \eqref{highsnrcdf} into  \eqref{opofq}, we can obtain the asymptotic OP of user $q$ as
\begin{align}\label{asympopqoptimal}
    OP_q^{\infty}(\gamma_{\rm th})&=F_{ h_q}^{\infty}\left(\frac{\gamma_{\rm th}^2+2\gamma_{\rm th}}{\overline{\gamma}Q_q}\right)\notag\\
    &=\frac{m_q^{m_q}d_{j_q=0}(\gamma_{\rm th}^2+2\gamma_{\rm th})}{2\sigma^2_qQ_q\Gamma(m_q)}\overline{\gamma}^{-1},
\end{align}
which shows that the diversity order of user $q$ is 1.

\subsection{Ergodic Capacity}
 Exploiting \eqref{rp} and \eqref{rq}, we can get $C_{\rm erg}$ as  
\begin{align}\label{aftermani}
    C_{\rm erg} = &\underbrace{{\mathbb{E}}\left[\log_2\left(1+\overline{\gamma}Q_ph_p+\sqrt{1+\overline{\gamma}Q_q h_q}\right)\right]}_{I_8}\notag\\
    &-\underbrace{{\mathbb{E}}\left[\log_2\left(1+1\Big/\sqrt{1+\overline{\gamma}Q_q h_q}\right)\right]}_{I_9}.
\end{align}

In the following, we will calculate $I_8$ and $I_9$, separately.
\subsubsection{Result for $I_8$}
Resorting to the Taylor's expansion of $\log_{2}(1+x)$ with the mean of $x$\cite{da2009capacity}, $I_8$ in \eqref{aftermani} can be approximated as 
\begin{align}
    I_8\approx e_0\cdot\left[\ln(1+{\mathbb{E}}\left[I_{10}\right])-\frac{{\mathbb{E}}\left[I_{10}^2\right]-E^2[I_{10}]}{2(1+{\mathbb{E}}\left[I_{10}\right])^2}\right],
    \label{I1}
\end{align}
where $I_{10}=\sqrt{1+\overline{\gamma}Q_q h_q}+\overline{\gamma}Q_p h_p$ and $e_0=\log_2e$.

As $ h_p$ and $ h_q$ are independent, we can obtain
\begin{align}
    {\mathbb{E}}\left[I_{10}\right]={\mathbb{E}}\left[\sqrt{1+\overline{\gamma}Q_q h_q}\right]+{\mathbb{E}}\left[\overline{\gamma}Q_p h_p\right]
    \label{EI_10}
\end{align}
and
\begin{align}
    {\mathbb{E}}\left[I_{10}^2\right]=&1+{\mathbb{E}}\left[\overline{\gamma}Q_q h_q\right]+{\mathbb{E}}\left[\overline{\gamma}^2Q_p^2h_p^2\right]\notag\\
    &+2{\mathbb{E}}\left[\sqrt{1+\overline{\gamma}Q_q h_q}\right]{\mathbb{E}}\left[\overline{\gamma}Q_p h_p\right].
\label{EI_102}
\end{align}

To evaluate \eqref{EI_10} and \eqref{EI_102}, we need to calculate two terms: ${\mathbb{E}}\left[\left(\overline{\gamma}Q_t h_t\right)^n\right]$ $(t\in\{p,q\})$ and ${\mathbb{E}}\left[\sqrt{1+\overline{\gamma}Q_q h_q}\right]$.

\paragraph{The calculation of ${\mathbb{E}}\left[\left(\overline{\gamma}Q_t h_t\right)^n\right]$ $(t\in\{p,q\})$}
Here, we first start with ${\mathbb{E}}\left[\left(\overline{\gamma}Q_p h_p\right)^n\right]$. Using \eqref{PDF} and \cite[Eq. (3.351-3)]{gradshteyn2007}, one can achieve

\begin{align}
    {\mathbb{E}}\left[\left(\overline{\gamma}Q_ph_p\right)^n\right]&=\overline{\gamma}^nQ_p^n\int_{0}^{\infty}x^nf_{ h_p}(x)\mathrm{d}x\notag\\
    &=\overline{\gamma}^nQ_p^n\sum_{j_p=0}^{\infty}\frac{H_p}{j_p!(2\sigma_p^2)^{j_p+1}}\notag\\
    & ~~~\times \int_{0}^{\infty}{x^{n+j_p}}{}\exp{\left(-\frac{x}{2\sigma_p^2}\right)}\mathrm{d}x\notag\\
    &=\alpha^{-n}\sum_{j_p=0}^{\infty}\frac{H_p}{j_p!}(j_p+n)!.
\label{n_power}
\end{align}

As $ h_p$ and $ h_q$ follow a same distribution, we can easily have 
\begin{align}
      {\mathbb{E}}\left[(\overline{\gamma}Q_q h_q)^n\right]=\beta^{-n}\sum_{j_q=0}^{\infty}\frac{H_q}{j_q!}(j_q+n)!.
\label{expectofy}
\end{align}

\paragraph{The calculation of ${\mathbb{E}}\left[\sqrt{1+\overline{\gamma}Q_q h_q}\right]$} Using \eqref{PDF} and \cite[Eq. (2.3.6-9)]{prudnikov1986integrals}, it deduces 

\begin{align}
        {\mathbb{E}}\left[\sqrt{1+\overline{\gamma}Q_qh_q}\right]&=\sum_{j_q=0}^{\infty}\frac{H_q}{j_q!(2\sigma_q^2)^{j_q+1}}\notag\\
        &~~~~\times \int_{0}^{\infty}y^{j_q}\sqrt{1+\overline{\gamma}Q_qy}\exp{\left(-\frac{y}{2\sigma_q^2}\right)}\mathrm{d}x\notag\\
        &=\sum_{j_q=0}^{\infty}H_q\beta^{j_q+1}\Psi(j_q+1,j_q+2.5;\beta),
\label{expectofsqrty}
\end{align}
where $\Psi(\cdot)$ is the confluent hypergeometric function.

Substituting \eqref{n_power} ($n=1$) and \eqref{expectofsqrty} into \eqref{EI_10}, we have
\begin{align}
    {\mathbb{E}}\left[I_{10}\right]&=\sum_{j_q=0}^{\infty}H_q\beta^{j_q+1}\Psi(j_q+1,j_q+2.5;\beta)\notag\\
    &~~~~+\alpha^{-1}\sum_{j_p=0}^{\infty}H_p(j_p+1).
\label{I3}
\end{align}

Substituting \eqref{n_power} ($n=2$), \eqref{expectofy} ($n=1$) and \eqref{expectofsqrty} into \eqref{EI_102}, ${\mathbb{E}}\left[I_{10}^2\right]$ can be given in \eqref{I32}, shown on the top of next page.

\begin{figure*}
    \begin{align}\label{I32}
        {\mathbb{E}}\left[I_{10}^2\right]=&\alpha^{-2}\sum_{j_p=0}^{\infty}H_p(j_p+1)(j_p+2)+1+\beta^{-1}\sum_{j_q=0}^{\infty}H_q(j_q+1)    +2\alpha^{-1}\sum_{j_p=0}^{\infty}H_p(j_p+1)  \sum_{j_q=0}^{\infty}H_q\beta^{j_q+1}\Psi(j_q+1,j_q+2.5;\beta)
    \end{align}
    \rule{18cm}{0.01cm}
\end{figure*}

Inserting \eqref{I3} and \eqref{I32} into \eqref{I1}, we will get the closed-form expression for $I_8$ as
\begin{align}\label{I1closedap}
 I_8\approx &\log_2\left(1+W+l_p\right)-\frac{e_0}{2}\cdot\left(1+W+{l_p}\right)^{-2}\notag\\
 &\times\Bigg[\alpha^{-1}l_p(j_p+2)+1+{l_q}-W^2-{l^2_p}\Bigg],
\end{align}
where $W=\sum\limits_{j_q=0}^{\infty}H_q\beta^{j_q+1}\Psi(j_q+1,j_q+2.5;\beta)$, $l_p=\alpha^{-1}\sum\limits_{j_p=0}^{\infty}H_p(j_p+1)$, and $l_q=\beta^{-1}\sum\limits_{j_q=0}^{\infty}H_q(j_q+1)$.

\subsubsection{Lower Bound of $I_8$}
As $f(x_1,x_2)=\log_2\big(1+\exp{(x_1)}+\exp{(x_2)}\big)$ is a convex function with respect to $x_1$ and $x_2$ \cite{zhong2012capacity}. Using Jensen's inequality, the lower bound of $I_8$ can be written as 
\begin{align}\label{lowerbound1}
    &I_8\geq\log_2\bigg(1+\exp{\left({\mathbb{E}}\left[\ln{\left(\sqrt{1+\overline{\gamma }Q_q h_q}\right)}\right]\right)}\notag\\
    &~~~~~~+\exp{\Big({\mathbb{E}}\left[\ln{\left(\overline{\gamma }Q_p h_p\right)}\right]\Big)}\bigg).
\end{align}

According to \cite[Eq. (11)]{adamchik1990algorithm}, $\ln\left(\sqrt{1+\overline{\gamma}Q_qh_q}\right)$ can be represented as
    $\ln(\sqrt{1+\overline{\gamma}Q_q h_q})=\frac{1}{2}\MeijerG*{1}{2}{2}{2}{1,1}{1,0}{\overline{\gamma}Q_q h_q}$.

Utilizing \eqref{PDF} and \cite[Eq. (2.6.2)]{mathai1978}, one can have

\begin{align}
    &{\mathbb{E}}\left[\ln\left({\sqrt{1+\overline{\gamma}Q_q h_q}}\right)\right]=\sum_{j_q=0}^{\infty}\frac{H_q}{2j_q!(2\sigma_q^2)^{j_q+1}}\notag\\
    &~~~~~~~~~~~~~~~~~~~~~~\times\int_{0}^{\infty}\exp{\left(-\frac{y}{2\sigma_q^2}\right)}y^{j_q}\MeijerG*{1}{0}{0}{1}{-}{0}{\frac{y}{2\sigma_q^2}}\mathrm{d}y\notag\\
    &~~~~~~~~~~~~~~~~~~~~~~=\frac{1}{2}\sum_{j_q=0}^{\infty}\frac{H_q}{j_q!}\MeijerG*{1}{3}{3}{2}{-j_q,1,1}{1,0}{\frac{1}{\beta}}.
\label{lnsqrty}
\end{align}

Employing \eqref{PDF} and \cite[Eq. (4.352-1)]{gradshteyn2007}, we can get 

\begin{align}
    {\mathbb{E}}\left[\ln{\left(\overline{\gamma}Q_p h_p\right)}\right]&=\ln{(\overline{\gamma}Q_p)}+{\mathbb{E}}\left[\ln{(h_p)}\right]\notag\\
    &=\ln{(\overline{\gamma}Q_p)}+\sum_{j_p=0}^{\infty}\frac{H_p}{j_p!(2\sigma_p^2)^{j_p+1}}\notag\\
    &~~~\times \int_{0}^{\infty}{x^{j_p}}\ln{(x)}{}e^{-\frac{x}{2\sigma_p^2}}\mathrm{d}x\notag\\
    &=-\ln{\alpha}+\sum_{j_p=0}^{\infty}H_p\psi(j_p+1),
\label{lnx}
\end{align}
where $\psi(\cdot)$ is the Euler psi function.

Substituting \eqref{lnsqrty} and \eqref{lnx} into \eqref{lowerbound1}, the lower bound of $I_8$ can be expressed as
\begin{align}
    I_8 \geq &\log_2\Bigg(\exp{\bigg(\frac{1}{2}\sum_{j_q=0}^{\infty}\frac{H_q}{j_q!}\MeijerG*{1}{3}{3}{2}{1,2,-j_q}{2,2}{2\overline{\gamma}Q_q\sigma_q^2}\bigg)}\notag\\
    &+\alpha^{-1}\exp{\bigg(\sum_{j_p=0}^{\infty}H_p\psi(j_p+1)\bigg)}+1\Bigg).
\label{lowerbound}
\end{align}

\subsubsection{Upper Bound of $I_8$}
As $f(x_1,x_2)=\log_2(1+x_1+x_2)$ is a concave function with respect to $x_1$ and $x_2$, by using Jensen's inequality, \eqref{EI_10}, and \eqref{I3}, we can obtain the upper bound of $I_8$ as
\begin{align}
    I_8&\leq\log_2\left(1+{\mathbb{E}}\left[\overline{\gamma}Q_p h_p\right]+{\mathbb{E}}\left[\sqrt{1+\overline{\gamma}Q_q h_q}\right]\right)\notag\\
    &=\log_2\left(1+{\mathbb{E}}\left[I_{10}\right]\right)\notag\\
    &=\log_2\bigg(1+\alpha^{-1}\sum_{j_p=0}^{\infty}H_p(j_p+1)\notag\\
    &~~~~+\sum_{j_q=0}^{\infty}H_q\beta^{j_q+1}\Psi(j_q+1,j_q+2.5;\beta)\bigg).
\label{upperbound}
\end{align}

\subsubsection{Result for $I_9$}
Substituting \eqref{PDF} into $I_9$ in \eqref{aftermani}, we can get

\begin{align}
        I_9=e_0\sum_{j_q=0}^{\infty}\frac{H_q}{j_q!(2\sigma_q^2)^{j_q+1}}\cdot I_{11},
\label{I2int1}
\end{align}
where \(I_{11}=\int_0^{\infty}y^{j_q}\ln{\left(1+(1+\overline{\gamma}Q_qy)^{-0.5}\right)}\exp{\left(-\frac{y}{2\sigma_q^2}\right)}\mathrm{d}y\).

Let $t=(1+\overline{\gamma}Q_qy)^{-0.5}$, then $y=\frac{1}{\overline{\gamma}Q_q}\left(\frac{1}{t^2}-1\right)$. $I_{11}$ in \eqref{I2int1} can be written as
\begin{align}
        &I_{11}=\frac{2\exp{(\beta)}}{(\overline{\gamma}Q_q)^{j_q+1}}\notag\\
        &~~~~~~\times \underbrace{\int_0^{1}\frac{(1-t^2)^{j_q}}{t^{2j_q+3}}\ln{(1+t)}\exp{\left(-\frac{\beta}{t^2}\right)}\mathrm{d}t}_{I_{12}}.
\label{I2int2}
\end{align}

Similar to \eqref{OPp6}, by setting $z=2t-1$ and using Chebyshev-Gauss  quadrature in the first case, we can obtain
\begin{align}
    I_{12}&=4\int_{-1}^{1}\frac{\left((z+3)(1-z)\right)^{j_q}}{(z+1)^{2j_q+3}}\ln{\left(\frac{z+3}{2}\right)}\notag\\
    &~~~~\times \exp{\left(-\frac{4\beta}{(z+1)^2}\right)}\mathrm{d}z\notag\\
    &=4\sum\limits_{k=1}^{J}\frac{\pi}{J}\sqrt{1-\phi_k^2}\frac{\left((\phi_k+3)(1-\phi_k)\right)^{j_q}}{(\phi_k+1)^{2j_q+3}}\notag\\
    &~~~~\times \ln{\left(\frac{\phi_k+3}{2}\right)}\exp{\left(-\frac{4\beta}{(\phi_k+1)^2}\right)},
\label{I2int3}
\end{align}
where $\phi_k=\cos\left(\frac{2k-1}{2J}\pi\right)$.

Using \eqref{I2int1}, \eqref{I2int2} and \eqref{I2int3}, $I_9$ can be expressed in \eqref{I_9}, shown on the top of next page.
\begin{figure*}
    \begin{align}
        I_9=&8e_0\cdot \exp{(\beta)}\sum_{j_q=0}^{\infty}\frac{H_q\beta^{j_q+1}}{j_q!} \sum\limits_{k=1}^{J}\frac{\pi}{J}\sqrt{1-\phi_k^2} \frac{\left((\phi_k+3)(1-\phi_k)\right)^{j_q}}{(\phi_k+1)^{2j_q+3}}\ln{\left(\frac{\phi_k+3}{2}\right)}\exp{\left(-\frac{4\beta}{(\phi_k+1)^2}\right)}
    \label{I_9}
    \end{align}
    \rule{18cm}{0.01cm}
\end{figure*}

Then, subtracting \eqref{I_9} separately from \eqref{I1closedap}, \eqref{lowerbound} and \eqref{upperbound}, we can get the closed-form expression, lower and upper bounds of $C_{\rm erg}$.

\section{Discussions on GPA and OPA schemes }\label{discusionGPAOPA}
\subsection{The OP of User $p$}\label{comparisonopp}
\eqref{oppgeneral} and \eqref{OPp11} show that user $q$'s OP under GPA and OPA schemes is the probabilities of $h_p\leq\frac{\gamma_{\rm th}}{a \overline{\gamma} Q_p}$ and $h_p\leq \frac{\gamma_{\rm th}\left(\sqrt{1+\overline{\gamma}Q_qh_q}+1\right)}{\overline{\gamma}Q_p}$, respectively.

When the OP under GPA scheme is greater than that under OPA scheme,  $\frac{\gamma_{\rm th}}{a \overline{\gamma} Q_p}> \frac{\gamma_{\rm th}\left(\sqrt{1+\overline{\gamma}Q_qh_q}+1\right)}{\overline{\gamma}Q_p}$ should be satisfied.

Thus, we can obtain
\begin{align}\label{oppdisc}
    \overline{\gamma}< \frac{1}{Q_qh_qa}\cdot\left(\frac{1}{a}-2\right).
\end{align}

Thus, it is easy to obtain that the OP under GPA scheme outperforms that under OPA scheme when $\overline{\gamma}> \frac{1}{Q_qh_qa}\cdot\left(\frac{1}{a}-2\right)$.

If $\overline{\gamma}\gg0$,  \eqref{asympoppgeneral} and \eqref{highsnrcdfp} show that $ OP_p^{\infty}(\gamma_{\rm th})\propto\overline{\gamma}^{-1}$ under GPA scheme and $ OP_p^{\infty}(\gamma_{\rm th})\propto\overline{\gamma}^{-1/2}$ under OPA scheme, respectively, which means that the OP under GPA scheme will decrease faster than that under OPA scheme when $\overline{\gamma}$ increases. In other words, GPA scheme outperforms OPA scheme in terms of OP in large $\overline{\gamma}$ region.

Therefore, some remarks can be drawn for the outage performance of user $p$ under the two considered power allocation schemes as follows: 

1) The relationship between the OP of user $p$ under these two schemes is determined by $\overline{\gamma}$ and independent of $\gamma_{\rm th}$, as suggested by \eqref{oppdisc}; 

2) When $\overline{\gamma}$ is small, the outage performance under GPA scheme is worse than that under OPA scheme, but not $vice$ $versa$ in large $\overline{\gamma}$ region.

\subsection{The OP of User $q$}\label{comparisonopq}
\eqref{opqgeneral} and \eqref{opofq} show that probabilities of $h_q\leq\frac{\gamma_{\rm th} }{ (1-a-a\gamma_{\rm th}) \overline{\gamma}Q_q}$ and $h_q\leq\frac{\gamma_{\rm th}^2+2\gamma_{\rm th}}{\overline{\gamma}Q_q}$ are the outage performance of user $q$ under GPA and OPA schemes, respectively.

If the outage performance under GPA scheme is worse than that under OPA scheme, $\frac{\gamma_{\rm th} }{ (1-a-a\gamma_{\rm th}) \overline{\gamma}Q_q}>\frac{\gamma_{\rm th}^2+2\gamma_{\rm th}}{\overline{\gamma}Q_q}$ with $1-a-a\gamma_{\rm th}>0$ should be satisfied. Then, we can acquire 
\begin{align}\label{rth11}
    \gamma_{\rm th}>\frac{1}{a}-2.
\end{align}

Similarly, when $\gamma_{\rm th}<\frac{1}{a}-2$, the outage performance under GPA scheme outperforms that under OPA scheme.

When $1-a-a\gamma_{\rm th}\leq 0$ which satisfies \eqref{rth11}, \eqref{opqgeneral} shows that the OP under GPA scheme is always equal to 1 which is greater than or equal to that under OPA scheme. 

When $\overline{\gamma}\gg0$ and $1-a-a\gamma_{\rm th}> 0$, \eqref{aympopqgeneral} and \eqref{asympopqoptimal} indicate that $ OP_q^{\infty}(\gamma_{\rm th})\propto\overline{\gamma}^{-1}$ under GPA scheme and $ OP_q^{\infty}(\gamma_{\rm th})\propto\overline{\gamma}^{-1}$ under OPA scheme, respectively, which means that $\overline{\gamma}$ exhibits the same effect on the outage performance under these two schemes. 

Finally, some remarks can be concluded for the outage performance of user $q$ under the two considered power allocation schemes as follows: 

1) In large $\overline{\gamma}$ region, $\overline{\gamma}$ exhibits the same impact on the OP under these two schemes;

2) The relationship between the OP of user $q$ under these two schemes is determined by $\gamma_{\rm th}$ and has nothing to do with $\overline{\gamma}$, as suggested by \eqref{rth11}; 

3) When $\gamma_{\rm th}$ is large, the outage performance under OPA scheme outperforms that under GPA scheme, but not $vice$ $versa$ in small $\gamma_{\rm th}$ region.

\subsection{Discussions on EC}
Obviously, one can easily obtain that the EC under OPA scheme is always better than that under GPA scheme since the purpose of OPA scheme is to achieve the optimal EC for the considered system, the rigorous proof of which is presented in Appendix I.

\section{Numerical Results}\label{simulations}

In this section, we conduct extensive numerical experiments to: 1) Compare the sum rates of two users under TDMA, GPA and OPA schemes; 2) Investigate the impacts of FTR channel parameters on outage and capacity performance under GPA and OPA schemes; 3) Validate our derived closed-form expressions; 4) Compare the outage and capacity performance in single-input single-output (SISO) and MIMO systems.

\subsection{The Comparison of Sum Rates under GPA, OPA, and TDMA Schemes}
\begin{figure}[!htb]
\centering
\includegraphics[width=3.1 in]{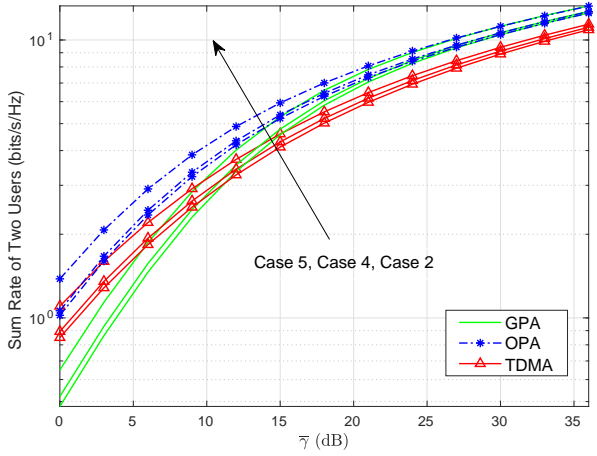}
    \caption{Sum rates of two users under GPA, OPA, and TDMA schemes for case 2, case 4, and case 5 in Table. \ref{tab:parameter} (in Appendix I) with $a=0.2$ for GPA scheme and $Q_p/Q_q=10$} 
    \label{fig:sumratecompare}
\end{figure}
Fig. \ref{fig:sumratecompare} shows the sum rates of the considered system under GPA, OPA, and TDMA schemes with three sets of FTR parameters and $Q_p/Q_q=10$. One can see that the transmitted average SNR $\overline{\gamma}$ shows a positive effect on sum rates. The sum rate under OPA scheme is always better than these under the two other schemes, which shows the correctness of our derivation of \eqref{rp} and \eqref{rq} in Appendix I. The sum rate under TDMA scheme is greater than that under GPA scheme when $\overline{\gamma}$ is small and the opposite conclusion can be drawn when $\overline{\gamma}$ is large. At high $\overline{\gamma}$, the sum rates under OPA and GPA are very close.

\subsection{The Impact of FTR Channel Parameters}
In this subsection, we explore the impacts of two users' FTR channel parameters on the OP and EC of considered systems. Observing the PDF and CDF of the FTR channel in \eqref{PDF} and \eqref{CDF}, one can get that $K_p, m_p, \Delta_p, \sigma_p$ are the four main parameters. For convenience, we set $\kappa_t=[K_t,m_t,\Delta_t,\sigma_t]$ ($t\in\{p,q\}$) and $a=0.2$ for all the simulations under GPA scheme.
\begin{figure}[!htb]
    \centering
    \includegraphics[width=3.1 in]{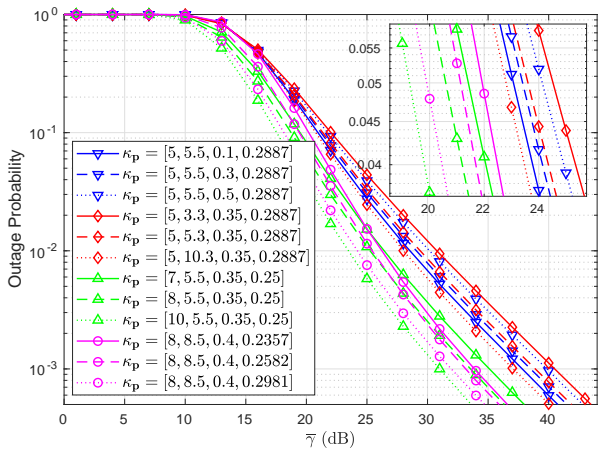}
    \caption{The OP of user $p$ with different FTR channel parameters under GPA scheme with $Q_p=1.5$ and $\gamma_{\rm th}=10$}
    \label{fig:OP_p_FTR_general}
\end{figure}

\begin{figure}[!htb]
    \centering
    \includegraphics[width=3.1 in]{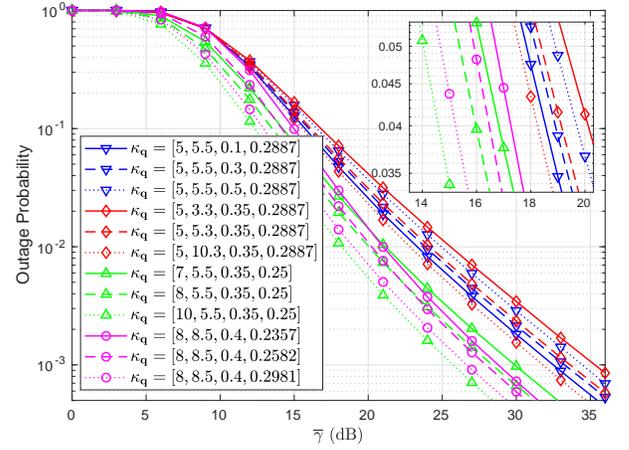}
    \caption{The OP of user $q$ with different FTR channel parameters under GPA scheme with $Q_q=0.5$ and $\gamma_{\rm th}=2$}
    \label{fig:OP_q_FTR_general}
\end{figure}

\eqref{snrp} and \eqref{snrq} shows that the SINR of the two users under GPA scheme is only affected by their own channel fading. Figs. \ref{fig:OP_p_FTR_general} and \ref{fig:OP_q_FTR_general} depict the impacts of the four main FTR channel parameters on the OP of users $p$ and $q$ under GPA scheme. The two users' OP increases as $\Delta_t$ ($t\in\{p,q\}$) increases or the three other parameters decrease. 

\begin{figure}[!htb]
    \centering
    \includegraphics[width=3.1 in]{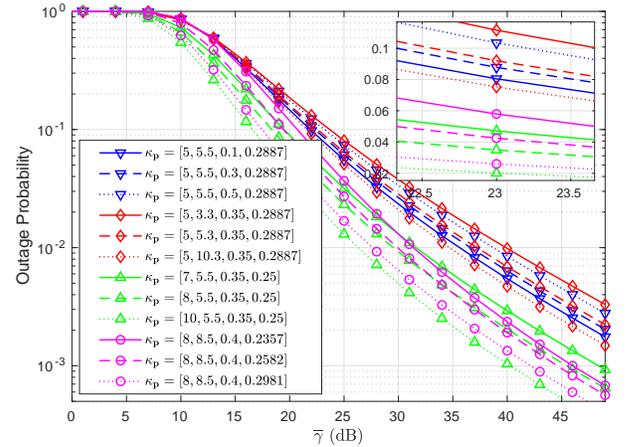}
    \caption{The OP of user $p$ with different FTR channel parameters of user $p$ under OPA scheme with $Q_p=1.5$ and $\gamma_{\rm th}=10$}
    \label{fig:OP_p_FTR_optimal_p}
\end{figure}

\begin{figure}[!htb]
    \centering
    \includegraphics[width=3.1 in]{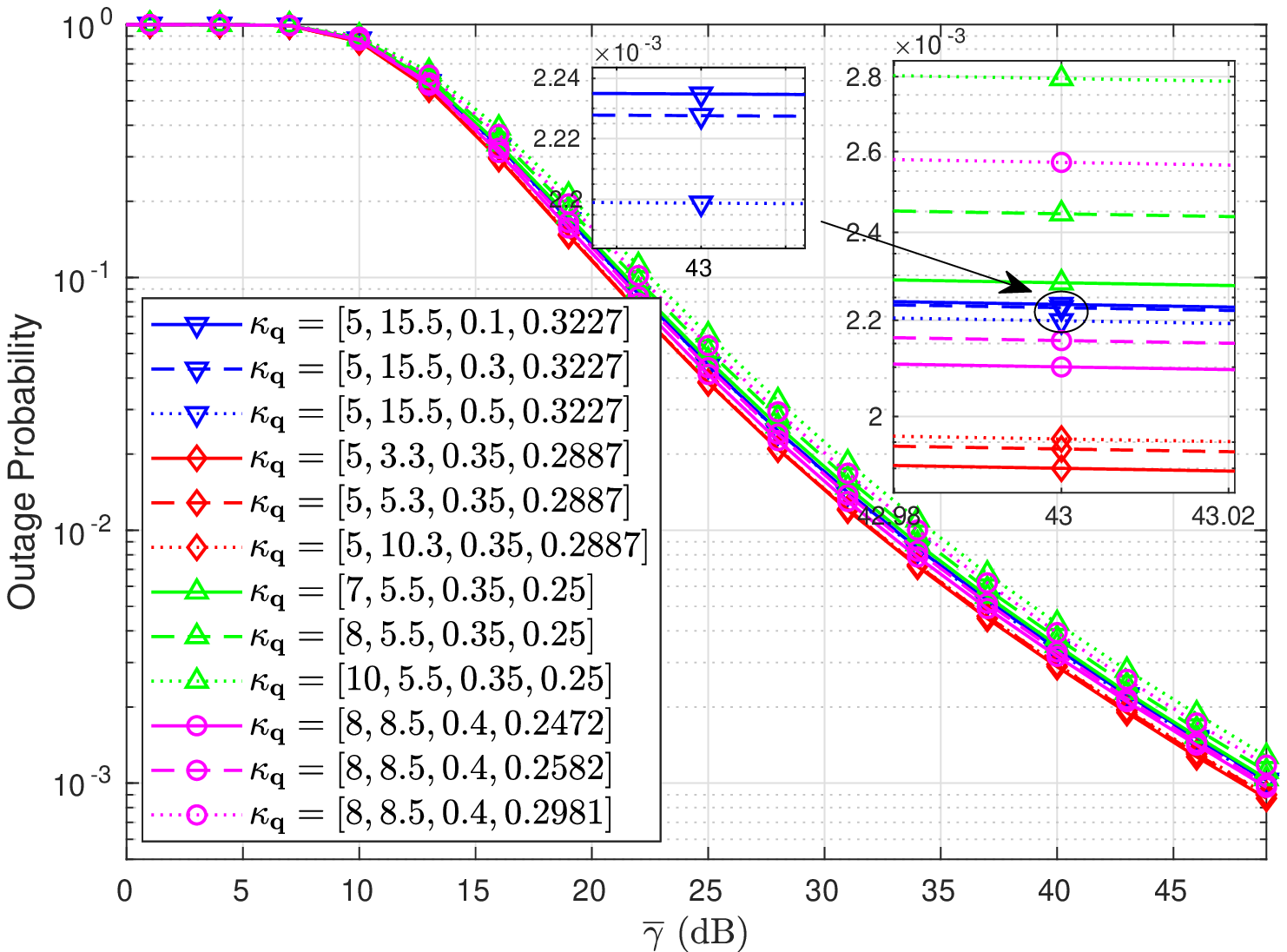}
    \caption{The OP of user $p$ with different FTR channel parameters of user $q$ under OPA scheme with $Q_p=1.5$ and $\gamma_{\rm th}=10$}
    \label{fig:OP_p_FTR_optimal_q}
\end{figure}

\begin{figure}[!htb]
    \centering
    \includegraphics[width=3.1 in]{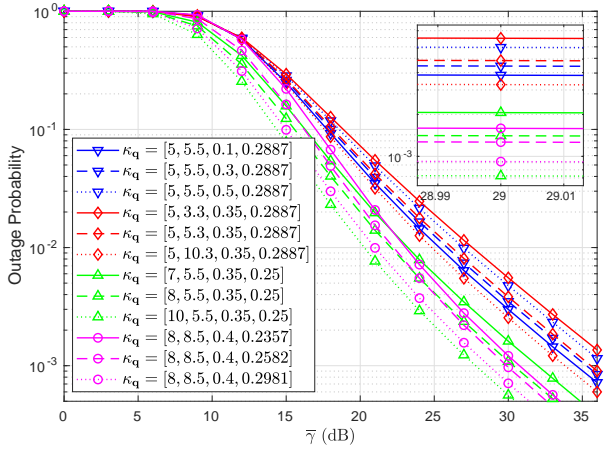}
    \caption{The OP of user $q$ with different FTR channel parameters under OPA scheme with $Q_q=0.5$ and $\gamma_{\rm th}=2$}
    \label{fig:OP_q_FTR_optimal}
\end{figure}

\eqref{rp} and \eqref{rq} present that the SINR of user $p$ under OPA scheme is determined by these two users' channel fading power gains $h_p$ and $h_q$ while user $q$'s SINR only relies on its own $h_q$. Figs \ref{fig:OP_p_FTR_optimal_p} and \ref{fig:OP_p_FTR_optimal_q} demonstrate the impacts of the two users' channel parameters on user $q$'s outage performance under OPA scheme. User $p$'s OP degrades as $\Delta_p, K_q, m_q, \text{ and } \sigma_q$ increase but decreases when $K_p, m_p, \sigma_p \text{ and }\Delta_q$ rise. Fig. \ref{fig:OP_q_FTR_optimal} depicts the impact of user $q$'s channel parameters on its own OP which gets worse with increasing $\Delta_q$ or decreasing the three other parameters.

\begin{figure}[!htb]
    \centering
    \includegraphics[width=3.1 in]{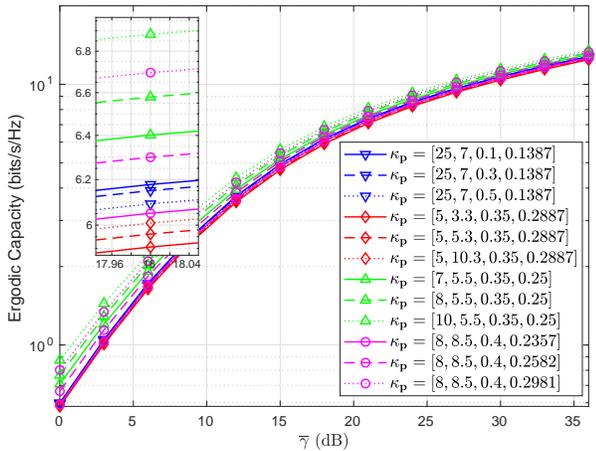}
    \caption{The EC with different FTR channel parameters of user $p$ under GPA scheme with  $Q_p=2$, $Q_q=0.1$, $K_q=5$, $m_q=15.5$, $\Delta_q=0.5$, and $\sigma_q=0.3162$}
    \label{fig:EC_p_FTR_general}
\end{figure}
\begin{figure}[!htb]
    \centering
    \includegraphics[width=3.1 in]{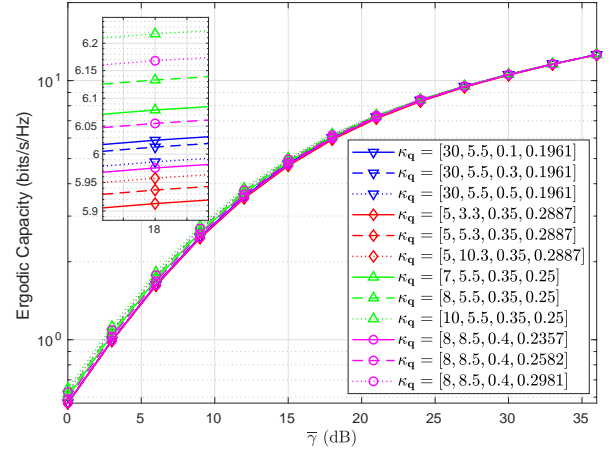}
    \caption{The EC with different FTR channel parameters of user $q$  under GPA scheme where  $Q_q=0.1$, $Q_p=2$, $K_p=8$, $m_p=5.5$, $\Delta_p=0.35$, and $\sigma_p=0.2357$}
    \label{fig:EC_q_FTR_general}
\end{figure}

Figs. \ref{fig:EC_p_FTR_general} and \ref{fig:EC_q_FTR_general} show the impact of FTR channel parameters of one user under GPA scheme on the EC of the considered system with a fixed channel of the other user. Both these two users' channels exhibit the same impacts on the capacity performance which deteriorates with increasing $\Delta_t$ and is enhanced with rising $K_t, m_t \text{ and } \sigma_t$ ($t\in\{p,q\}$).

\begin{figure}[!htb]
    \centering
    \includegraphics[width=3.1 in]{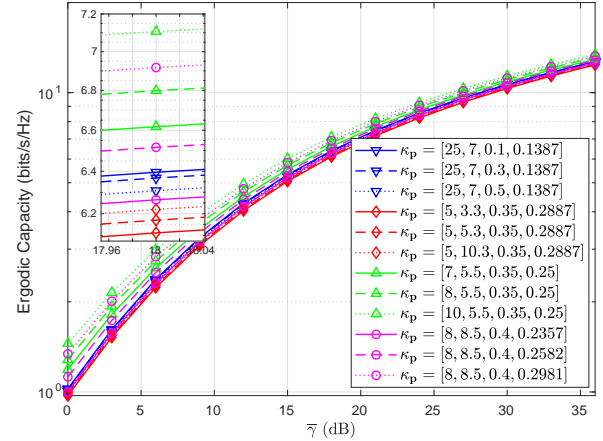}
    \caption{The EC with different FTR channel parameters of user $p$ under OPA scheme with $Q_p=2$, $Q_q=0.1$, $K_q=5$, $m_q=15.5$, $\Delta_q=0.5$, and $\sigma_q=0.3162$}
    \label{fig:EC_p_FTR_optimal}
\end{figure}
\begin{figure}[!htb]
    \centering
    \includegraphics[width=3.1 in]{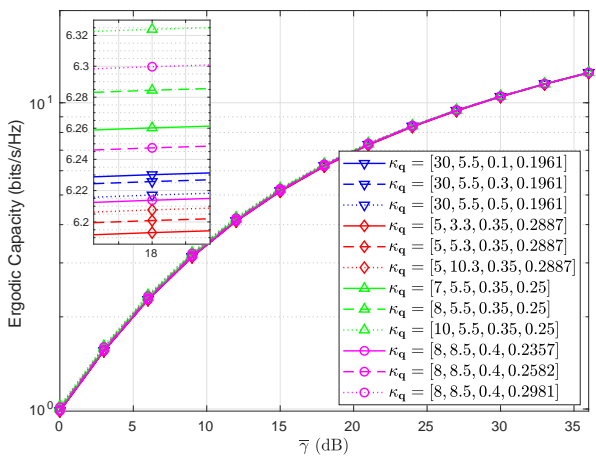}
    \caption{The EC with different FTR channel parameters of user $q$ under OPA scheme with $Q_q=0.1$, $Q_p=2$, $K_p=8$, $m_p=5.5$, $\Delta_p=0.35$, and $\sigma_p=0.2357$}
    \label{fig:EC_q_FTR_optimal}
\end{figure}

Figs. \ref{fig:EC_p_FTR_optimal} and \ref{fig:EC_q_FTR_optimal} present how the EC of the considered system is affected by FTR channel parameters of one user under OPA scheme while the channel of the other user is fixed. The EC degrades with increasing $\Delta_t$ or decreasing $K_t, m_t \text{ and } \sigma_t$ ($t\in\{p,q\}$).

In summary, $\Delta_t$ ($t\in\{p,q\}$) exhibits a negative effect on user $t$'s OP and the EC of the considered system while $K_t, m_t \text{ and } \sigma_t$ show positive effects under both GPA and OPA schemes. Under OPA scheme, large $\Delta_q$ leads to enhanced user $p$'s OP while large $K_q, m_q, \text{ and } \sigma_q$ cause the degraded outage performance.

\subsection{Validation of Derived Expressions and Comparisons between SISO and MIMO Cases}
\begin{figure}[!htb]
\centering
\subfigure[GPA scheme]{
\begin{minipage}[t]{1\linewidth}
\centering
\includegraphics[width=3.1 in]{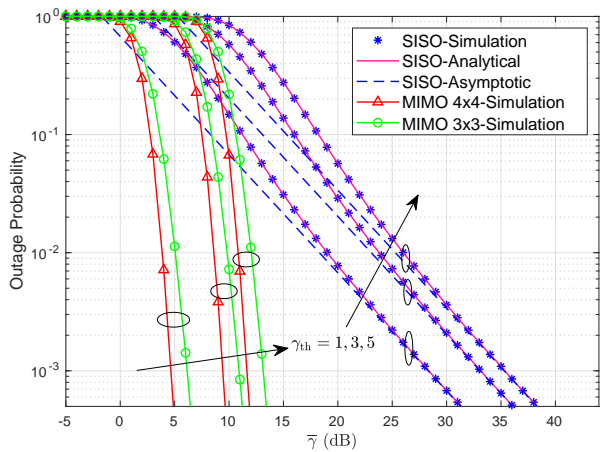}
\label{opp_general}
\end{minipage}%
}%

\subfigure[OPA scheme]{
\begin{minipage}[t]{1\linewidth}
\centering
\includegraphics[width=3.1 in]{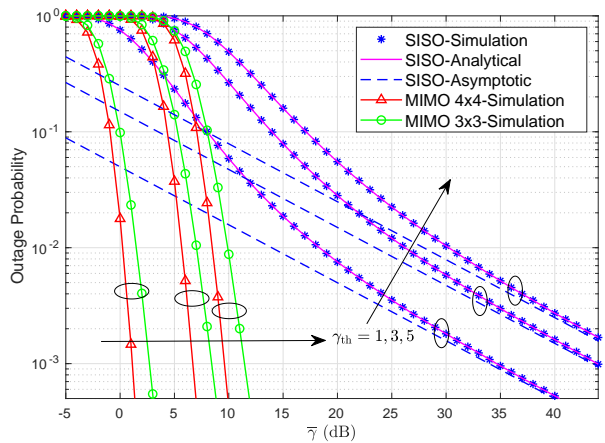}
\label{opp_optimal}
\end{minipage}%
}
\centering
\caption{The OP of user $p$ versus $\overline{\gamma}$ with $K_p=5$, $m_p=10.8$, $\Delta_p=0.5$, $\sigma_p=0.2887$, $Q_p=1.5$ $K_q=10$, $m_q=5.5$, $\Delta_q=0.35$, $\sigma_q=0.2132$, and $Q_q=0.15$}
\label{fig:opp}
\end{figure}

\begin{figure}[!htb]
\centering
\subfigure[GPA scheme]{
\begin{minipage}[t]{1\linewidth}
\centering
\includegraphics[width=3.1 in]{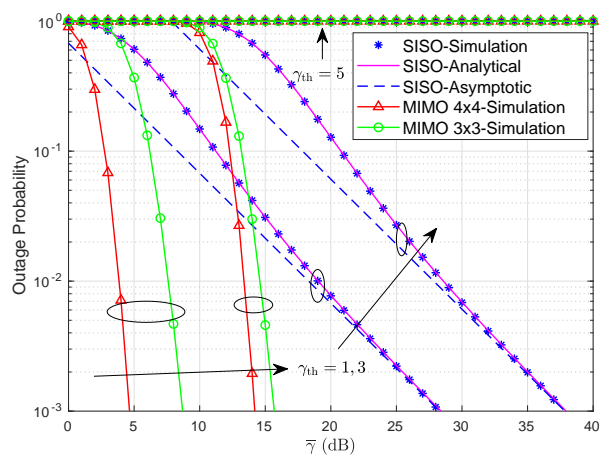}
\label{Opq_general}
\end{minipage}%
}%

\subfigure[OPA scheme]{
\begin{minipage}[t]{1\linewidth}
\centering
\includegraphics[width=3.1 in]{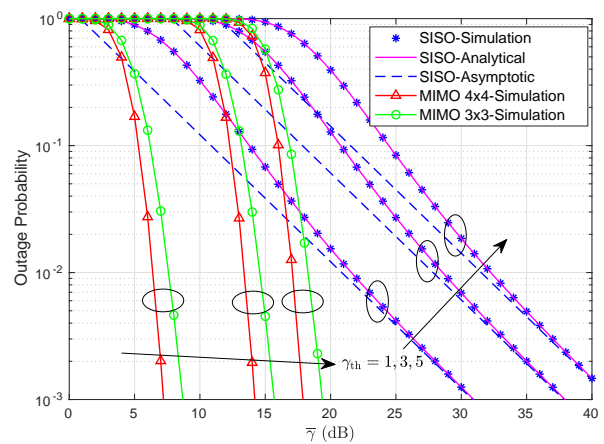}
\label{opq_optimal}
\end{minipage}%
}
\centering
\caption{The OP of user $q$ versus $\overline{\gamma}$ with $K_q=5$, $m_q=0.5$, $\Delta_q=10.8$, $\sigma_q=0.2887$, and $Q_q=0.5$}
\label{fig:opq}
\end{figure}

In previous subsections, the considered system in SISO cases were studied. In this subsection, we conduct extensive numerical experiments to validate these derived expressions and compare the considered system in both SISO and MIMO cases to show the impacts of the diversity gains achieved via MIMO. Numerical results are obtained by MATLAB and PYTHON programming to validate the theoretical analysis. In the MIMO $t\times r$ system with $t$ transmit antennas and $r$ receiving antennas, selection combining method is utilized to choose the channel link with the best channel condition. During computing the PDF and CDF of $ h_t$ ($t\in(p,q)$), we use the finite terms instead of the infinite summation terms, which is shown feasible in \cite{zhang2017new},\cite{zeng2018physical}. In this section, we adopt the first 80 terms of the summation which can also achieve a very high precision\cite{zhao2019secure}, and $10^7$ channel state realizations are generated to conduct Monte-Carlo simulations.

 Figs. \ref{fig:opp} and \ref{fig:opq} present the OP of user $p$ and $q$ versus $\overline{\gamma}$ in SISO and MIMO cases under GPA and OPA schemes. The SISO's analysis curves under these two schemes match their simulation ones well, which confirms the correctness of our derived closed-form expression. Generally, when $\overline{\gamma}$ increases, OP decreases, which means increasing $\overline{\gamma}$ can improve the outage performance. Given a same $\overline{\gamma}$, a high $\gamma_{\rm th}$ leads to the high OP, which means the degradation of outage performance. At the same time, the analysis curves converge to the asymptotic ones in the high SNR region, which verifies our asymptotic expression. Notably, when $a=0.2$ and $\gamma_{\rm th}=5$ satisfy $a_{\rm th}=1-a-a\gamma_{\rm th}\leq0$ in \eqref{opqgeneral}, simulation, analytical, and asymptotic OP of user $q$ under GPA scheme are all equal to 1 at any $\overline{\gamma}$ which can be seen from Fig. \ref{Opq_general}. Moreover, we can also easily see that the slopes of the asymptotic curves in Figs. \ref{opp_general}, \ref{Opq_general}, \ref{opp_optimal}, and \ref{opq_optimal} agree well with the behaviors of simulation and analysis results in high $\overline{\gamma}$ region, which implies the correctness of the derived diversity order. From Fig. \ref{fig:opp}, we can see that the outage performance of user $p$ under GPA scheme is worse than that under OPA scheme when $\overline{\gamma}$ is small and the opposite conclusion can be obtained when $\overline{\gamma}$ is large, which verifies the remarks in Section \ref{comparisonopp}. From Fig. \ref{fig:opq}, we can observe that the OP of user $q$ under GPA scheme is smaller than that under OPA scheme at when $\gamma_{\rm th}$ is small and the former is larger than the latter when $\gamma_{\rm th}$ is large, which shows the correctness of the remarks in Section \ref{comparisonopq}.

\begin{figure}[!htb]
\centering
\subfigure[GPA scheme]{
\begin{minipage}[t]{1\linewidth}
\centering
\includegraphics[width=3.1 in]{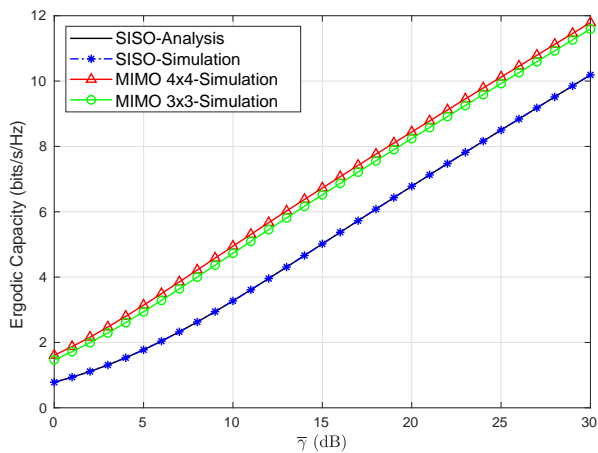}
\label{ec_general}
\end{minipage}%
}%

\subfigure[OPA scheme]{
\begin{minipage}[t]{1\linewidth}
\centering
\includegraphics[width = 3.1 in]{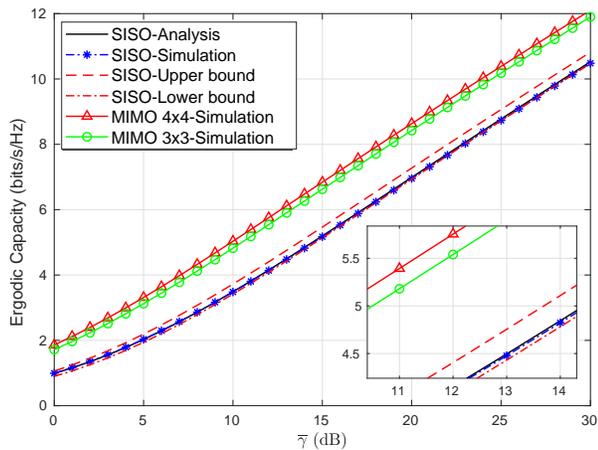}
\label{ec_optimal}
\end{minipage}%
}
\centering
\caption{EC versus $\overline{\gamma}$ with $K_p=8$, $m_p=5.5$, $\Delta_p=0.35$,$\sigma_p=0.2357$, $Q_p=2$ $K_q=5$, $m_q=15.5$, $\Delta_q=0.5$, $\sigma_q=0.3162$, and $Q_q=0.1$}
\label{fig:ec}
\end{figure}

Fig. \ref{ec_general} manifests the simulation and analysis results of EC versus the average SNR $\overline{\gamma}$ under GPA scheme. Fig. \ref{ec_optimal} shows the simulation and analysis results, and the lower and upper bound of EC versus the average SNR $\overline{\gamma}$ under OPA schemes. As shown in these two figures, EC increases while $\overline{\gamma}$ arises, which means that increasing $\overline{\gamma}$ leads to improved capacity performance. We can also see that the curves of the analysis match very well with those of their Monte-Carlo simulations even though we applied approximation during the derivation in SISO case under OPA scheme. In Fig. \ref{ec_optimal}, the gap between the upper and lower bounds is very narrow and they bind the simulation and analysis curves tightly. We can observe from these two figures that the EC under OPA scheme is larger than that under GPA, which verifies our derivation of OPA scheme again.

Finally, one can also easily observe from Figs. 11-13 that the EC and OP for MIMO cases outperform those for SISO and more antennas can provide better outage and capacity performance, because of the diversity gain brought by multiple transmissions and receiving under MIMO cases.

\section{Conclusion}\label{conclusion}
In this paper, we have studied the OP and EC of NOMA over FTR channels in mmWave communication under GPA and OPA schemes. Under the former scheme, the closed-form expressions for OP and EC, as well as asymptotic expressions for OP were derived. Under the other scheme, the closed-form and asymptotic expressions for OP, and the closed-form, upper bound and lower bound expressions for the EC were also derived. Monte-Carlo simulation was conducted to verify these proposed expressions. We also investigated the effects of the number of antennas on EC and OP.

Some useful remarks are reached as follows:
\begin{itemize}
\item The relationship between the OP of user $p$ under GPA and OPA schemes is determined by $\overline{\gamma}$ and independent of $\gamma_{\rm th}$, as suggested by \eqref{oppdisc};
\item When $\overline{\gamma}$ is small, the outage performance of user $p$ under GPA scheme is worse than that under OPA scheme, but not $vice$ $versa$ in large $\overline{\gamma}$ region. 
\item  In large $\overline{\gamma}$ region, $\overline{\gamma}$ exhibits the same impact on the OP of user $q$ under GPA and OPA schemes;
\item The relationship between the OP of user $q$ under GPA and OPA schemes is determined by $\gamma_{\rm th}$ and has nothing to do with $\overline{\gamma}$, as suggested by \eqref{rth11};
\item When $\gamma_{\rm th}$ is large, the outage performance of user $q$ under OPA scheme outperforms that under GPA scheme, but not $vice$ $versa$ in small $\gamma_{\rm th}$ region;
\item OPA scheme exhibits a better performance of the sum rate than GPA and TDMA.
\item Generally, FTR channel parameter $\Delta_t$ ($t\in\{p,q\}$) has a negative effect on the OP and EC of user $t$ while $K_t, m_t \text{ and } \sigma_t$ show positive effects. Notably, $\Delta_q$ exhibits positive impacts, and the three other parameters show negative effects on user $p$'s OP under OPA scheme.
\item Under GPA scheme, the diversity orders of user $p$ and $q$ are both 1. Under OPA scheme, the diversity order of user $p$ is 0.5 and that of user $q$ is 1.  
\item More antennas can provide better outage and capacity performance, because of the achieved diversity gains. 
\end{itemize}

\section*{Appendix I: The derivation of SINRs under OPA scheme }

Under TDMA scheme, the capacity of user $p$ and $q$ in the same time slots can be written as
\begin{align}\label{tdmap}
    R_p^{TDMA}=0.5\log_2(1+\overline{\gamma}Q_ph_p)
\end{align}
and
\begin{align}\label{tdmaq}
    R_q^{TDMA}=0.5\log_2(1+\overline{\gamma}Q_qh_q).
\end{align}

According to \eqref{sumrate}, to guarantee that both of these two users can be regarded as a primary user, their capacities under NOMA scheme are always better than those under TDMA scheme.

For user $p$, we have
\begin{align}
    \log_2(1+a \overline{\gamma}Q_p h_p)\ge 0.5\log_2(1+\overline{\gamma}Q_ph_p),\notag
\end{align}

so that,

\begin{align}
     a \ge\frac{1}{\sqrt{1+\overline{\gamma}Q_ph_p}+1}.
\end{align}

Similarly, for user $q$, we get
\begin{align}
    \log_2\left(1+\frac{(1-\alpha)\overline{\gamma}Q_qh_q}{a \overline{\gamma}Q_qh_q+1}\right)\ge 0.5\log_2(1+\overline{\gamma}Q_qh_q),\notag
\end{align}

that is
\begin{align}
     \overline{\gamma}Q_qh_q-\sqrt{1+\overline{\gamma}Q_qh_q}+1 \ge a \overline{\gamma}Q_qh_q\sqrt{1+\overline{\gamma}Q_qh_q}, \notag
\end{align}

hence
\begin{align}
     \frac{1}{\sqrt{1+\overline{\gamma}Q_qh_q}+1} \ge a.
\end{align}

So we can get the desired range of $a$ as $\frac{1}{\sqrt{1+\overline{\gamma}Q_ph_p}+1}\leq a \leq \frac{1}{\sqrt{1+\overline{\gamma}Q_qh_q}+1}$.

The first derivative of $R_{sum}$ in \eqref{sumrate} with respect to $a$ is \begin{align}\label{derivative}
    \frac{\partial R_{sum}}{\partial\alpha}&=\frac{1}{\ln{2}}\cdot\left(\frac{P_sQ_ph_p}{\alpha P_sQ_ph_p+N_0}-\frac{P_sQ_qh_q}{\alpha P_sQ_qh_q+N_0}\right)\notag\\
    &=\frac{1}{\ln{2}}\cdot \frac{N_0P_s(Q_ph_p-Q_qh_q)}{(\alpha P_sQ_ph_p+N_0)(\alpha P_sQ_qh_q+N_0)}.
\end{align}

\begin{figure}[!htb]
     \centering
    \includegraphics[width=3.1 in]{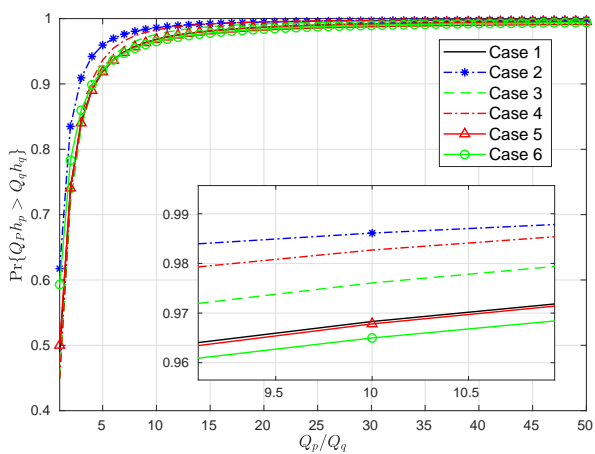}
    \caption{Probability of $Q_ph_p>Q_qh_q$ for 6 cases in Table. \ref{tab:parameter}}
    \centering
    \label{fig:QpQq}
\end{figure}

\renewcommand{\arraystretch}{1.5}
\begin{table}[!htb]
    \centering
    \caption{Parameter Settings of FTR model}
    \begin{tabular}{ c| c| c| c| c|c}
    \hline\hline
     Case Number & User $t$ & $m_t$ & $K_t$ & $\Delta_t$ & $\sigma_t$\\
     \hline
    \multirow{2}{*}{Case 1} & $p$ & 10.8 & 5 & 0.5 & 0.2887 \\ 
    & $q$ & 5.5 & 10 & 0.35 & 0.2132 \\  \hline
    \multirow{2}{*}{Case 2} & $p$ & 5.5 & 8 & 0.35 & 0.2981 \\ 
    & $q$ & 15.5 & 5 & 0.5 & 0.3162 \\  \hline
    \multirow{2}{*}{Case 3} & $p$ & 5.5 & 8 & 0.1 & 0.2357 \\ 
    & $q$ & 3.3 & 10 & 0.4 & 0.2335 \\  \hline
    \multirow{2}{*}{Case 4} & $p$ & 15.5 & 8 & 0.35 & 0.2357 \\ 
    & $q$ & 3.3 & 15 & 0.4 & 0.1936 \\  \hline
    \multirow{2}{*}{Case 5} & $p$ & 10.8 & 5 & 0.5 & 0.2887 \\ 
    & $q$ & 10.8 & 5 & 0.5 & 0.2887 \\  \hline
    \multirow{2}{*}{Case 6} & $p$ & 3.5 & 5 & 0.5 & 0.3162 \\ 
    & $q$ & 3.5 & 5 & 0.5 & 0.2739 \\ 
     \hline
     \hline
    \end{tabular}
    \label{tab:parameter}
\end{table}

We randomly select 6 sets of parameters from the parameter settings in \cite{zhao2019secure, zhao2018different,zhang2017new,zheng2019wireless} shown in Table. \ref{tab:parameter}.

Under the setting of these parameters, the probabilities of $Q_ph_p>Q_qh_q$ versus $Q_p/Q_q$ are depicted in Fig. \ref{fig:QpQq}. One can see that when $Q_p \gg Q_q$ ($Q_p/Q_q>10$), ${\rm Pr}\{Q_ph_p>Q_qh_q\}>0.96$ which means that $Q_ph_p>Q_qh_q$ is equivalent to $Q_p\gg Q_q$. Therefore, we can get that $\frac{\partial R_{sum}}{\partial\alpha}$ in \eqref{derivative} is strictly positive, which implies that $R_{sum}$ is monotonically increasing in the range of $\frac{1}{\sqrt{1+\overline{\gamma}Q_ph_p}+1}\leq a \leq \frac{1}{\sqrt{1+\overline{\gamma}Q_qh_q}+1}$. The maximum sum rate can be achieved at $a_{opt}=\frac{1}{\sqrt{1+\overline{\gamma}Q_qh_q}+1}$. Substituting $a_{opt}$ into \eqref{snrp} and \eqref{snrq}, we can obtain new SINRs of users $p$ and $q$ shown in \eqref{rp} and \eqref{rq}.

\section*{Appendix II: The feasibility of SIC at user $p$ under OPA scheme }
\begin{figure}[!htb]
     \centering
    \includegraphics[width=3.1 in]{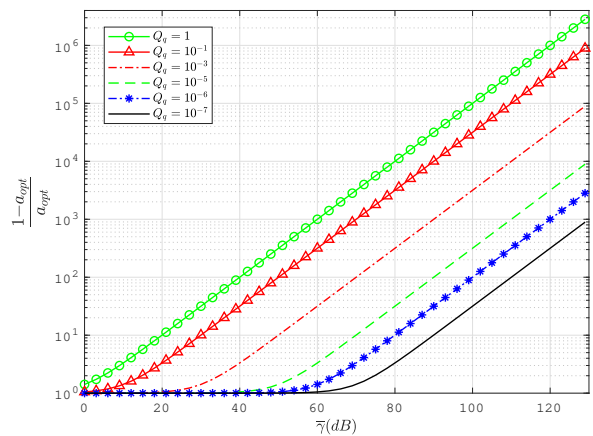}
    \caption{$\frac{1-a_{opt}}{a_{opt}}$ versus $\overline{\gamma}$ with different $Q_q$ and $\mathbb{E}\{h_q\}=1$}
    \centering
    \label{fig:aopt_Qq}
\end{figure}

\begin{figure}[!htb]
     \centering
    \includegraphics[width=3.1 in]{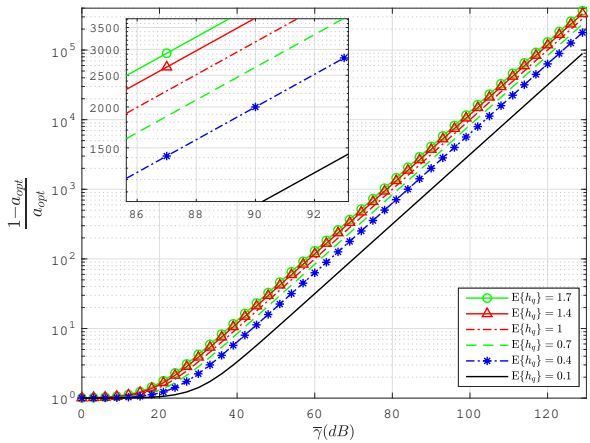}
    \caption{$\frac{1-a_{opt}}{a_{opt}}$ versus $\overline{\gamma}$ with different ${\mathbb{E}}\{h_q\}$ and $Q_q=0.01$}
    \centering
    \label{fig:aopt_Eh}
\end{figure}

{In this subsection, we will explain that the feasibility of the SIC at user $p$ under OPA scheme.}

{According to the key idea of the NOMA scheme, the received signals of the two users can be distinguished at user $p$ when the difference between their received power is large enough. Therefore, considering the fact that all received signals at user $p$ experience the same fading and path-loss, the SIC at user $p$ is feasible when the difference between their allocated transmit power is large enough.}

{Obviously, the larger the difference between the allocated transmit power of the two users is, the larger the ratio of the allocated transmit power of users $q$ and $p$, $\frac{1-a_{opt}}{a_{opt}}$, achieves. Then, a conclusion can be reached as: the larger $\frac{1-a_{opt}}{a_{opt}}$ is, the larger feasibility of SIC is.}

{In the following, some numerical results with practical settings of mmWave systems will be given to show the typical range of $\frac{1-a_{opt}}{a_{opt}}$, so as to illustrate the feasibility of the SIC  at user $p$ under OPA scheme. Fig. \ref{fig:aopt_Qq} shows $\frac{1-a_{opt}}{a_{opt}}$ versus $\overline{\gamma}$ with the expectation of $h_q$, ${\mathbb{E}}\{h_q\}=1$, and $Q_q=\{1,10^{-1},10^{-3},10^{-5},10^{-6},10^{-7}\}$. Fig. \ref{fig:aopt_Eh} presents $\frac{1-a_{opt}}{a_{opt}}$ versus $\overline{\gamma}$ with $Q_q=0.01$ and ${\mathbb{E}}\{h_p\}=\{1.7,1.4,1,0.7,0.4,0.1\}$.}

{In the following, we first demonstrate the reasonableness and practicality of the settings of $Q_q$ and $\overline{\gamma}$ adopted in Figs. \ref{fig:aopt_Qq} and \ref{fig:aopt_Eh}.}

{Let's consider an mmWave communication system operating at 28 GHz and the distance between the base station and user $q$ ranges from 10 to 200 meters. Thus, we can have the line-of-sight path-loss from 81 to 108 dB. The transmit antenna gain of the base station varies from 23 to 29 dBi \cite{huo20175g}. If the base station is equipped with massive MIMO antennas, larger transmit antenna gain as high as 36 dBi can be obtained \cite{busari2017millimeter}. The gain of mmWave receiver ranges from 14 to 45.5 dB\cite{bhattacharya2019system, pulipati2019design}. Therefore, $Q_q$, which contains the path loss, transmit and receive antenna gains, etc., can range from $10^{-7}$ to $1$. Hence, the setting of $Q_q$ considered in Figs. \ref{fig:aopt_Qq} and \ref{fig:aopt_Eh}, $Q_q=\{1,10^{-1},10^{-3},10^{-5},10^{-6},10^{-7}\}$, is reasonable.}

{Generally, the transmit power of mmWave systems varies from 16 to 40 dBm \cite{huo20175g,hemadeh2017millimeter,bhattacharya2019system}. If the bandwidth of the transmitted signal is from 0.5 to 2 GHz, the received power of the thermal noise will be from $-80$ to $-74$ dBm. So the range of $\overline{\gamma}$ normally ranges from 90 to 120 dB, which has been fully covered by the settings adopted in Figs. \ref{fig:aopt_Qq} and \ref{fig:aopt_Eh}.}

{As presented in Figs. \ref{fig:aopt_Qq} and \ref{fig:aopt_Eh}, one can clearly observe that $\frac{1-a_{opt}}{a_{opt}}$ is larger than 10 under numerical cases of $Q_q$ and ${\mathbb{E}}\{h_p\}$ when $\overline{\gamma}$ ranges from 90 to 120 dB.} 

{Hence, we can conclude that the SIC at user $p$ is feasible under OPA scheme. }

\bibliography{citation}
\end{document}